\documentclass[letterpaper]{article}

\usepackage[T1]{fontenc}
\usepackage{multicol}
\usepackage{hyperref}
\usepackage{wrapfig}
\usepackage{geometry}
\usepackage{setspace}

\usepackage[style=chem-acs, backend=biber]{biblatex}
\usepackage{graphicx}
\usepackage{float}
\newfloat{scheme}{htbp}{los}
\floatname{scheme}{Scheme}
\floatname{chart}{Chart}
\newfloat{graph}{htbp}{loh}

\usepackage{chemformula} % Formulas using \ch{}
\usepackage[version = 4]{mhchem} % Formulas using \ce{}
\usepackage{siunitx}
\usepackage{authblk}
\author[1]{Rahil Haria}
\author[1]{Noah Schnitzer}
\author[1,2]{T. Ben Britton}
\author[1]{Yaqi Li}
\author[3]{Tom J. P. Irons}
\author[1,4,5]{Sophia Linssen-Pitsaros}
\author[6,7]{Ella Banyas}
\author[1]{Geri Topore}
\author[1]{Annabel Hoyes}
\author[1]{Mariana Palos}
\author[6,7*]{Sinéad M. Griffin}
\author[3*]{Katherine Inzani}
\author[1*]{Michele Shelly Conroy}
\affil[1]{Department of Materials, Imperial College London, London, SW7 2AZ, United Kingdom}
\affil[2]{Department of Materials Engineering, UBC, Vancouver, British Columbia, Canada}
\affil[3]{School of Chemistry, University of Nottingham, Nottingham NG7 2RD, UK}
\affil[4]{Department of Physics and Astronomy, University College London, Gower Street, London WC1E 6BT, United Kingdom}
\affil[5]{London Centre for Nanotechnology, 17-19 Gordon Street, London WC1H 0AH, United Kingdom}
\affil[6]{Materials Sciences Division, Lawrence Berkeley National Laboratory, Berkeley, Berkeley, CA
94720, USA}
\affil[7]{Molecular Foundry, Lawrence Berkeley National Laboratory, Berkeley, Berkeley, CA 94720,
USA}

\title{Polar Topologies in a Ferroelastic Metal Membrane}
\date{*mconroy@imperial.ac.uk, katherine.inzani1@nottingham.ac.uk, sgriffin@lbl.gov}

\begin{document}

\maketitle

\begin{abstract}
\normalsize
Polar metals, materials in which electric polarisation and metallicity coexist, are exceptionally rare because itinerant electrons screen long-range dipoles and favour centrosymmetric structures. Engineering polar textures in a conducting magnet holds promise for reconfigurable spin-orbit coupling and magnetoelectric functionality. Here we show that releasing epitaxial SrRuO\textsubscript{3} films from their substrates drives a hierarchy of ferroelastic domain refinement from micrometre to nanometre length scales, and that this structural reorganisation spontaneously generates two distinct classes of emergent polar texture that are ubiquitous across the freestanding membrane. Using correlative microscopy from mesoscale electron channelling contrast imaging (ECCI) to  atomic-resolution scanning transmission electron microscopy (STEM), we demonstrate that electric polarisation emerges selectively at translation-inequivalent antiphase boundaries (APBs). At these boundaries multicomponent $a^-a^-c^+$ tilt field undergoes N\'{e}el-like interpolation that preserves the in-phase tilt component and amplifies roto-flexoelectric coupling, while translation-equivalent boundaries remain nonpolar. The N\'{e}el-like interpolation at hard APBs and Ising-like collapse of all tilt components at easy APBs is corroborated with ab initio calculations. While embedded $90^\circ$ ferroelastic walls provide an additional mechanistically distinct source of electric polarisation resulting in polar nanoclusters ($\sim$4~nm). These distinct nano-textures at $90^\circ$ walls from via elastic accommodation of strain mismatch between variants and rotostriction as the tilt field interpolates across the boundaries. These findings show that, in a membrane form, metal oxides provide a robust platform for hosting nanoscale ferroelastic domains that generate polar textures.
\end{abstract}

\onehalfspacing


\section*{Introduction}

Electric polarisation and metallicity are fundamentally contra-indicated properties, yet their coexistence is highly sought after: polarisation underpins piezoelectricity \cite{RN483}, magnetoelectric coupling \cite{doi:10.1126/science.1184028}, ferroelectricity \cite{doi:10.1126/science.1129564}, and spintronic functionality \cite{doi:10.1126/science.1065389}, functionalities that would be transformative if realised in a conducting system. Yet itinerant electrons screen static dipoles 
on the scale of the Thomas-Fermi length, suppressing long-range polar order and favouring centrosymmetric structures \cite{PhysRevLett.14.217, RN404, benedek2016ferroelectric}. Conventional mechanisms for polarisation, such as the second-order Jahn-Teller effect, require empty-valence-shell cations and favour insulating behaviour \cite{RN404, RN484}, making intrinsic polar metals exceedingly rare \cite{PhysRevLett.14.217, RN485, RN486}. Yet this incompatibility is not absolute: metallic screening suppresses homogeneous dipoles but does not forbid nanoscale polar displacements driven by spatial gradients of structural order parameters. Flexoelectricity - the linear coupling between polarisation and strain gradients that is symmetry allowed in all solids \cite{zubkoFlexoelectricEffectSolids2013, tagantsevPiezoelectricityFlexoelectricityCrystalline1986} - provides precisely such a route, generating polar displacements wherever structural inhomogeneities produce large strain gradients, even within a centrosymmetric metallic host.

SrRuO$_3$ is one of the most important electrode materials due to its high electrical conductivity and its ability to form atomically sharp epitaxial interfaces with perovskites \cite{eom1992single, KosterRevModPhys2012, nair2018synthesis, lichtensteiger2023nanoscale}. Recently, SrRuO$_3$ has attracted renewed attention in condensed matter physics following reports of magnetic phenomena such as the topological Hall effect \cite{Matsuno2016, qin2019emergence, meng2019observation}, chiral spin fluctuations \cite{wang2019spin}, and skyrmion phases \cite{wang2018ferroelectrically}. It has also been shown that strain gradients at the heterointerface between SrRuO$_3$ and SrTiO$_3$(111) can induce polar displacements \cite{RN415}, overcoming conventional screening to enable a flexoelectrically polarised metal. This establishes SrRuO$_3$ as a rare platform in which electronic polarisation, metallicity, and magnetism coexist and are intrinsically coupled \cite{RN404} thus potentially controlled locally via strain engineering. However, to date these polar phases have been confined to small strain gradient regions at heterointerfaces. 

Freestanding oxide membranes provide a powerful platform to amplify gradient-driven polar effects \cite{RN489, Paskiewicz2016, Ji2019, Dong2019, li2022electrostatically}. Releasing a film from its substrate eliminates long-range epitaxial constraints, allowing ferroelastic variants to refine, reorient, and reorganise across multiple length scales. The resulting nanodomain architectures concentrate strain gradients at structural defects: domain walls, antiphase boundaries, and ferroelastic junctions, creating conditions favourable for flexoelectric polarisation that are simply inaccessible in the substrate-clamped geometry. Freestanding SrRuO$_3$ nanomembranes have been demonstrated as a route to flexible oxide electronics and strain-relaxed lattice configurations \cite{Paskiewicz2016, Ji2019, Dong2019}. Zhou et al.~\cite{ZhouGiantPolar2025} reported polar Ru off-centring in freestanding SrRuO$_3$ films achieved through controlled strain relaxation. Recent work has shown that twisted bilayer stacking of metallic SrRuO$_3$ membranes can generate polar vortices through moiré-periodic flexoelectricity, demonstrating that gradient-driven polar textures can survive itinerant carrier screening in a ferromagnetic metal \cite{LunTwistronics2025}. Whether analogous polar textures can be stabilised intrinsically within a single freestanding membrane, without requiring deliberate geometric manipulation of stacked layers, remains an open question.

Here we show that freestanding SrRuO$_3$ membranes develop a hierarchy of emergent polar textures driven by nanoscale ferroelastic domain refinement. Using correlative electron microscopy across multiple length scales, from micrometre-scale ECCI to atomic-resolution STEM combined with Fourier-space strain analysis, we map the full ferroelastic domain architecture of both substrate-clamped and freestanding films and identify the symmetry-distinct boundary types responsible for polar order. We show that polarisation emerges selectively at translation-inequivalent antiphase boundaries, where the multicomponent $a^-a^-c^+$ tilt field undergoes Néel-like interpolation, preserving the $M_3^+$ component through the wall and amplifying roto-flexoelectric coupling precisely where tilt gradients are steepest, while translation-equivalent boundaries, where all tilt components undergo Ising-like collapse simultaneously, remain nonpolar. These results establish the Ising versus Néel interpolation character of ferroelastic antiphase boundaries, determined by the translational relationship between domains, as a design principle for engineering emergent polar-magnetic functionality in conducting oxide membranes. Embedded $90^\circ$ ferroelastic walls provide a second, mechanistically distinct source of polarisation, nucleating polar nanoclusters of characteristic size $\sim$4~nm through localised hydrostatic strain gradients arising from elastic accommodation of the deviatoric eigenstrain mismatch between variants and rotostriction as the tilt field interpolates across the boundary. The resulting polar displacement fields exhibit locally divergent and convergent vector configurations bearing resemblance to those found within topological polar textures such as vortices and skyrmions reported in polar oxide systems \cite{Das2019, Yadav2016}. These textures are stabilised here within a single freestanding conducting layer without twisted stacking geometries or dedicated polar heterointerfaces, representing to our knowledge the first observation of topology-seeded polar textures in a single freestanding metallic membrane.

\section*{Results}

\subsection*{Domain Mapping in Epitaxial and Freestanding SrRuO\textsubscript{3}}

\begin{figure}[htb!]
\centering
\includegraphics[width=\columnwidth]{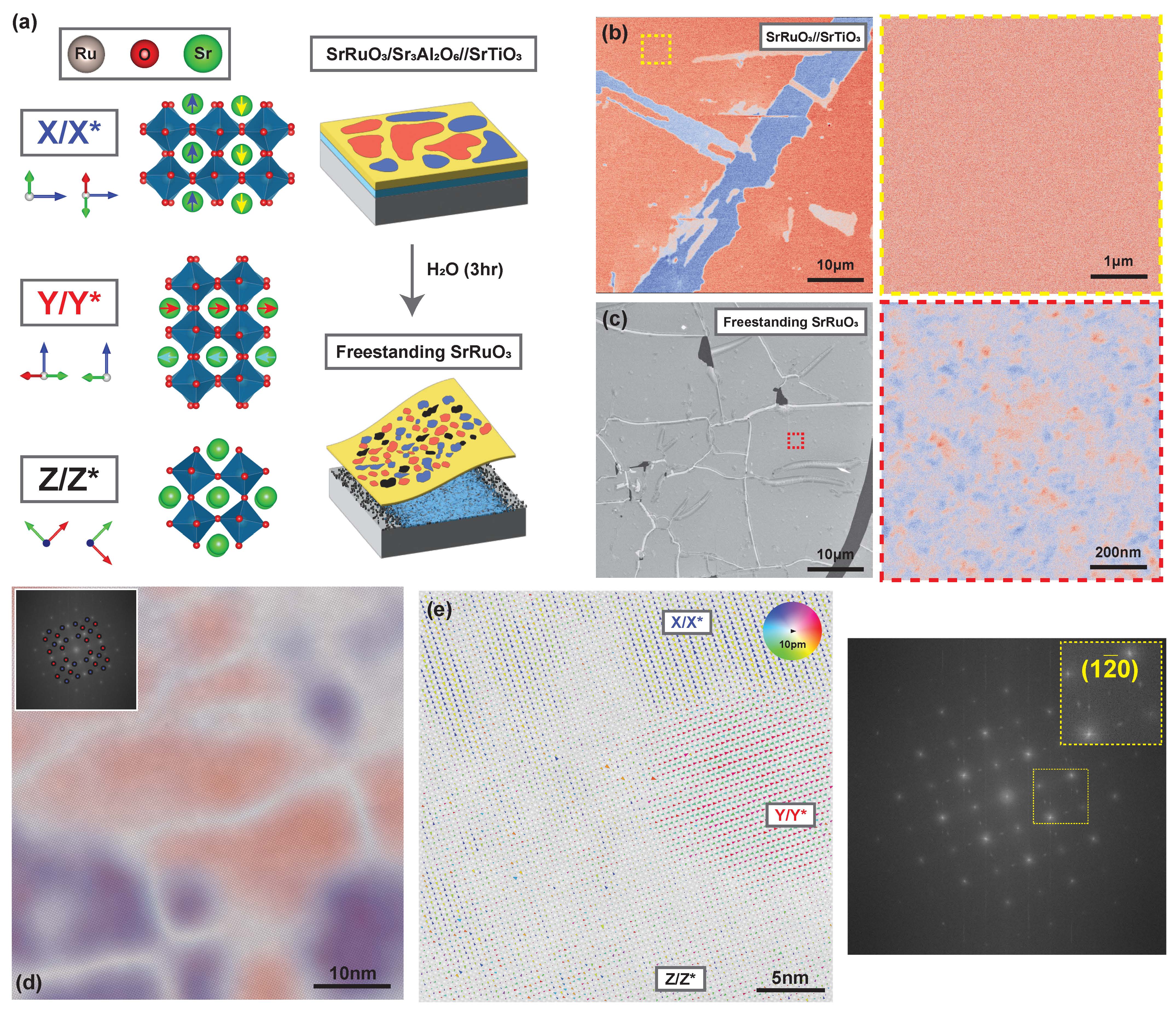}
\caption{
\textbf{Ferroelastic domain mapping in epitaxial and freestanding SrRuO\textsubscript{3}.}
\textbf{(a)} Six symmetry-equivalent orientation variants of SrRuO\textsubscript{3} (X/X$^{*}$, Y/Y$^{*}$, Z/Z$^{*}$) with crystallographic axes labelled ($a$ red, $b$ green, $c$ blue). Schematic of release from SrRuO\textsubscript{3}/Sr\textsubscript{3}Al$_2$O$_6$//SrTiO\textsubscript{3} to a freestanding membrane after H$_2$O etching, illustrating the transition from micrometre-scale X/X$^{*}$ and Y/Y$^{*}$ domains in the epitaxial film to a refined nanodomain mixture of X/X$^{*}$, Y/Y$^{*}$, and Z/Z$^{*}$ variants.
\textbf{(b)} ECCI of epitaxial SrRuO\textsubscript{3}/SrTiO\textsubscript{3}. Low magnification image reveals well-defined ferroelastic domains with lateral dimensions on the order of tens of micrometres; false colour is applied to regions of differing contrast to aid visualisation and is illustrative only. High magnification image of a single domain interior showing uniform contrast, confirming orientational homogeneity.
\textbf{(c)} ECCI of freestanding SrRuO\textsubscript{3}. Low magnification image showing large-scale surface wrinkling arising from mechanical relaxation of the released membrane. High magnification image revealing contrast variations on the scale of tens of nanometres, providing direct evidence of a dense nanodomain population; false colour is applied for visual clarity and does not constitute variant assignment.
\textbf{(d)} Annular Dark Field STEM (ADF-STEM) image of freestanding SrRuO\textsubscript{3} with Fourier amplitudes of X/X$^{*}$ (blue) and Y/Y$^{*}$ (red) overlaid. Inset: corresponding FFT.
\textbf{(e)} picometer-scale periodic lattice displacements (PLD) mapping of X/X$^{*}$, Y/Y$^{*}$, and Z/Z$^{*}$ regions in the freestanding film. Inset: FFT showing the $1\bar{2}0$ reflection characteristic of the Z/Z$^{*}$ orientation.
}
\label{fig: ECCI and PLD}
\end{figure}

To investigate the structural hierarchy of SrRuO\textsubscript{3} (SRO) membranes, we first establish the orientations of the bulk orthorhombic lattice as a reference (Fig.~\ref{fig: ECCI and PLD}a). Bulk SRO exhibits the conventional ferroelastic variants, X/X* and Y/Y*, corresponding to the (110)/(1$\bar{1}$0) orientations, with the distinction between X and Y made visible through the Sr sublattice staggering, highlighted atop the Sr atoms in the figure. The Z/Z* variant is oriented along (001), appearing cubic-like in projection, and exhibits no coherent Sr staggering along the out-of-plane direction. These bulk orientations provide a baseline for understanding the effects of epitaxial clamping and subsequent freestanding relaxation. Diffraction patterns for each variant are provided in Fig.~\ref{fig: Table_for_Stuctures_and_HKL}.

ECCI provides a non-destructive approach to mapping mesoscale ferroelastic domains in epitaxial thin 
films~\cite{pengFerroelasticWritingCrystal2025, pengElectronChannelingContrast2026}. 
ECCI is sensitive to small local lattice rotations because variations in the misorientation of crystal planes relative to the incident electron beam modulate the backscattered electron intensity \cite{Qaiser:yr5161}. 
In epitaxially clamped SRO, the X/X$^*$ and Y/Y$^*$ variants are 
misoriented with respect to the $(001)$ film plane by an angle set by 
the clamping condition, and aligning the [001] pole along the electron 
optic axis links contrast level directly to variant type, permitting 
crystallographic assignment \cite{pengFerroelasticWritingCrystal2025, 
pengElectronChannelingContrast2026}. At low magnification, 
Fig.~\ref{fig: ECCI and PLD}b reveals well-defined ferroelastic domains with lateral dimensions on the order of tens of micrometres, consistent with elastic compatibility predictions for substrate-clamped films \cite{pengFerroelasticWritingCrystal2025}; false colour is applied for visual clarity. At high magnification, the interior of a single domain is contrast-free, confirming orientational uniformity. In freestanding SRO membranes (Fig.~\ref{fig: ECCI and PLD}c), the substrate-defined reference frame is absent: contrast instead arises from local lattice rotations between adjacent nanodomains and from membrane curvature. 
At low magnification, large-scale surface wrinkling reflects mechanical relaxation of the released film; at high magnification, nanoscale contrast variations on the scale of tens of nanometres constitute direct evidence of a dense population of crystallographically distinct domains, though unambiguous variant assignment is no longer possible without the epitaxial reference frame.

To quantitatively characterise the nanodomain structure, periodic lattice displacement (PLD) mapping of the STEM data provides domain characterisation~\cite{RN474}. In PLD mapping, atomic positions are first refined to $\sim 2$~pm precision, and Sr sites are isolated by masking every alternate column. Fourier-space analysis is then used to selectively suppress structural modulations associated with staggered Sr displacements spanning two pseudocubic unit cells. Damping the corresponding Fourier components isolates X/X* (blue) and Y/Y* (red) variants. The Fourier amplitudes of those variants are overlaid on the ADF-STEM image in Fig.~\ref{fig: ECCI and PLD}d, revealing the ferroelastic nanodomains. Subtracting the damped reconstruction from the original lattice produces a real-space map of Sr displacements (Fig.~\ref{fig: ECCI and PLD}e). This approach reveals the structure of the ferroelastic nanodomains, including Z/Z* (001)-oriented regions identified by the characteristic $1\bar{2}0$ reflection in the FFT (Fig.~\ref{fig: ECCI and PLD}e), which exist alongside conventional X/X* and Y/Y* variants at length scales below the resolution of ECCI (Fig.~\ref{fig: ECCI and PLD}c). These STEM observations therefore confirm that the reduced ECCI contrast in freestanding membranes originates from extensive domain refinement into nanoscale ferroelastic variants.

The emergence of a dense ferroelastic nanodomain network in freestanding SRO membranes arises from the relaxation of epitaxial clamping and the associated boundary conditions. Upon release from the STO substrate, long-range elastic strain is relieved, but the film remains constrained by internal compatibility conditions, leading to the subdivision of larger domains into nanoscale variants. This hierarchical refinement minimises the total elastic energy by accommodating local lattice rotations and misfit stresses, resulting in a high density of X/X*, Y/Y*, and Z/Z* subdomains.

\subsection*{Ferroelastic Nanodomains and Polar Antiphase Boundaries}

\begin{figure}[hbt!]
\centering
\includegraphics[width=0.9\columnwidth]{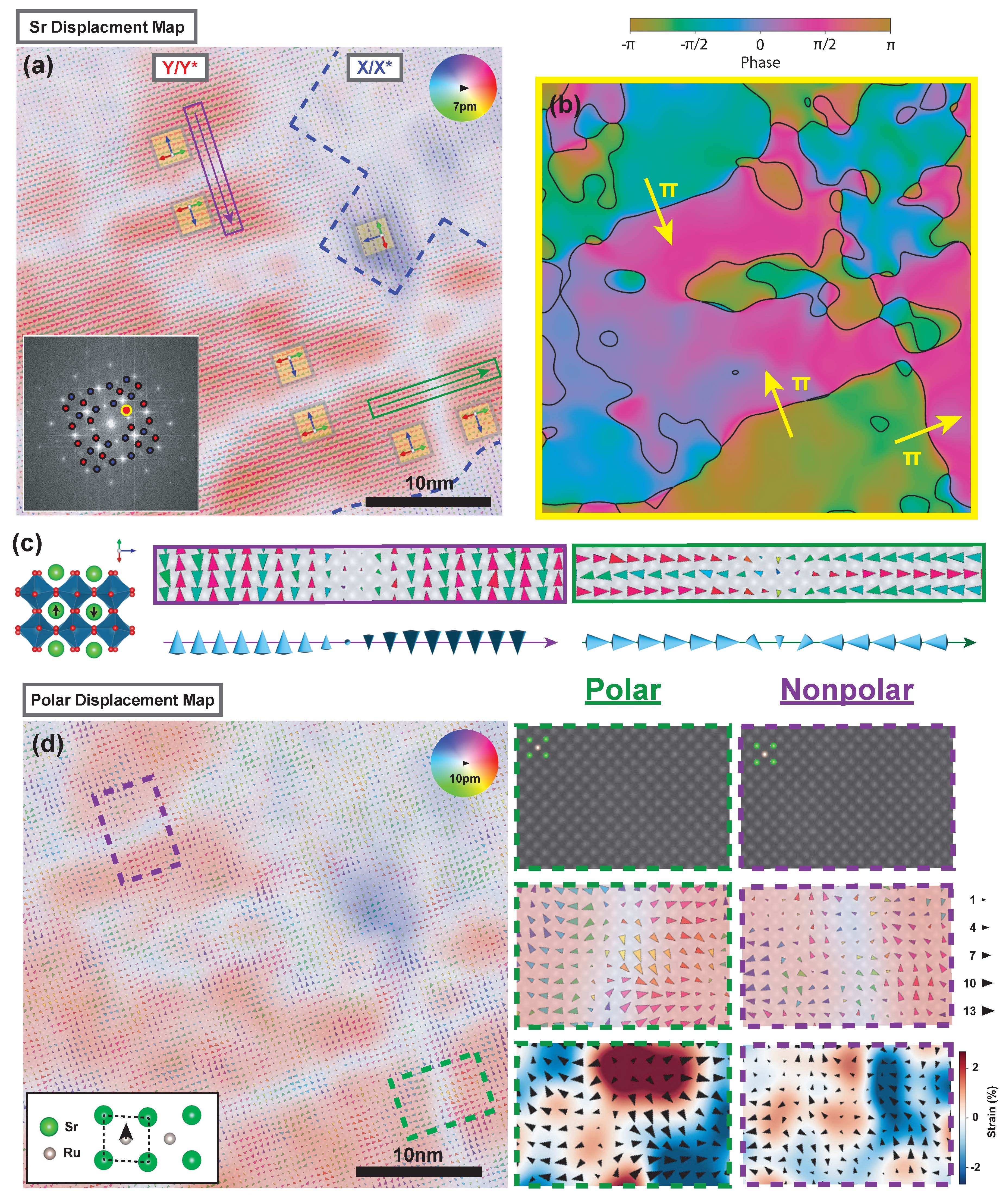}
\caption{\small 
\textbf{Ferroelastic nanodomain structure and antiphase boundary character in freestanding SrRuO\textsubscript{3}.} 
\textbf{(a)} Sr displacement map from PLD analysis with X/X$^*$ (blue) and Y/Y$^*$ (red) Fourier amplitudes overlaid. Purple and green arrows mark translation-equivalent (easy) and translation-inequivalent (hard) antiphase boundaries respectively.
\textbf{(b)} Fourier phase map of the $(11\bar{1})$ reflection showing discrete $180^\circ$ phase reversals at both boundaries.
\textbf{(c)} Magnified staggered A-site order parameter. The easy wall shows Ising-like interpolation ($|\mathbf{A}|$ suppressed, $\hat{\mathbf{A}}$ fixed); the hard wall shows N\'{e}el-like interpolation ($\hat{\mathbf{A}}$ rotates continuously, $|\mathbf{A}|$ weakly suppressed).
\textbf{(d)} Polar displacement map (Ru displacement within Sr cage) and phase lock-in strain maps with polarisation overlaid. No polar response at the easy wall (purple); polar displacements exceeding 13~pm directed away from the wall at the hard boundary (green), with the strain maps confirming a flexoelectric origin.
}
\label{fig: Long range Domain Mapping}
\end{figure}

Figure~\ref{fig: Long range Domain Mapping}a presents the Sr displacement map 
extracted from STEM PLD analysis, overlaid with the Fourier amplitudes of the X/X* (blue) and Y/Y* (red) ferroelastic variants. The inset shows the masked reciprocal-space peaks used to isolate the staggered Sr modulation. The reconstruction reveals a dense nanodomain network in the freestanding membrane.

Two representative $180^\circ$ domain walls are marked: a purple arrow indicating a translation-equivalent (easy) wall and a green arrow marking a translation-inequivalent (hard) wall.
Figure~\ref{fig: Long range Domain Mapping}b shows the Fourier phase map of the $(11\overline{1})$ reflection. Across both walls, the Y/Y* domains exhibit a discrete phase reversal of the staggered Sr modulation, confirming the presence of APBs. The phase discontinuity is identical at the purple and green arrows, indicating that both walls involve reversal of the structural modulation.

The distinction between easy and hard walls becomes evident at higher magnification (Fig.~\ref{fig: Long range Domain Mapping}c). To quantify the local wall character, we define a staggered A-site order parameter \[\mathbf{A}(\mathbf{r})=\langle \eta(\mathbf{r})\,\mathbf{u}_{\mathrm{Sr}}(\mathbf{r}) \rangle,\] where $\mathbf{u}_{\mathrm{Sr}}$ is the Sr displacement relative to the parent cubic position and $\eta=\pm1$ labels alternating Sr planes along $[001]_\mathrm{pc}$ corresponding to the two-sublattice staggering. The magnitude $|\mathbf{A}|$ measures the amplitude of the staggered distortion, while $\hat{\mathbf{A}}$ captures its orientation. The staggered Sr displacement accommodates the RuO$_6$ octahedral rotations, described by the primary tilt order parameter $\boldsymbol{\phi} = (\phi_x, \phi_y, \phi_z)$, where $(\phi_x, \phi_y)$ are the antiphase $R_4^+$ components and $\phi_z$ is the in-phase $M_3^+$ component. The RuO$_6$ octahedral rotations displace the A-site Sr ions from their cubic positions through rotostriction; $\mathbf{A}$ is therefore a secondary order parameter linearly coupled to $\boldsymbol{\phi}$, with $\mathbf{A} \propto \boldsymbol{\phi}$ to leading order. The experimentally observed behaviour of $\mathbf{A}$ across each boundary thus provides a direct experimental measure of the spatial evolution of the primary tilt field $\boldsymbol{\phi}$, which is the quantity that drives roto-flexoelectric coupling. At the easy wall (purple), $|\mathbf{A}|$ is strongly suppressed at the interface with negligible rotation of $\hat{\mathbf{A}}$, consistent with Ising-like interpolation on the order parameter: all components of $\boldsymbol{\phi}$ collapse toward zero together at the wall centre without rotating. At the hard wall (green), $\hat{\mathbf{A}}$ rotates continuously across the boundary with comparatively weaker amplitude suppression, indicating Néel-like interpolation: the in-plane components $(\phi_x, \phi_y)$ pass through zero and rotate while the out-of-plane component $\phi_z$ remains large throughout. The mechanistic consequences of this distinction for polar order are discussed in the following section.

The polar consequences are shown in Fig.~\ref{fig: Long range Domain Mapping}d, where an in-plane polarisation vector for each unit cell is calculated from the off-centring of each ruthenium relative to the four surrounding strontium atomic columns \cite{jia2007unit}. In the purple easy-wall region, no preferred polar direction is associated with the boundary. In the green hard-wall region, clear in-plane polar displacements exceeding 13~pm emerge, oriented away from the wall on both faces, extending over a total width of $\sim$8~nm above the experimental noise floor. Polarisation is therefore localised specifically at translation-inequivalent boundaries.

To understand the structural origin of this selectivity, we extract spatially resolved strain fields using Fourier phase lock-in analysis, relating local phase variations of selected FFT peaks to changes in lattice spacing along the corresponding crystallographic directions \cite{goodgeDisentanglingCoexistingStructural2022a}. Independent analysis along two orthogonal lattice vectors yields the full in-plane strain tensor, comprising the longitudinal components $\varepsilon_{xx}$ and $\varepsilon_{yy}$, measured along the in-plane pseudocubic $[100]_\mathrm{pc}$ and $[010]_\mathrm{pc}$ directions respectively, and the shear components $\varepsilon_{xy}$ and $\varepsilon_{yx}$; all four components are provided in Fig.~\ref{fig: Strain} of the Supplemental Information. From the longitudinal components, the in-plane hydrostatic strain,
\begin{equation}
\varepsilon_{\mathrm{hyd}} = \frac{\varepsilon_{xx} + \varepsilon_{yy}}{2},
\end{equation}
is computed, providing a spatially resolved map of local in-plane volumetric expansion and compression within the membrane. We note that this quantity is strictly a two-dimensional, projection-averaged measure of the dilatational strain, as ADF-STEM imaging provides no direct access to the out-of-plane strain component $\varepsilon_{zz}$; the term hydrostatic strain is used throughout as a shorthand for this in-plane dilatational quantity rather than implying full three-dimensional isotropy. These strain maps are overlaid with the polar displacement field in Fig.~\ref{fig: Long range Domain Mapping}d, enabling direct spatial comparison between local strain gradients and polar response at each boundary type.

At the hard (translation-inequivalent) boundary, the hydrostatic strain map reveals a sharp spatial variation tightly co-localised with the wall: tensile strain of $+2.2\%$ at the boundary interface drops abruptly to $-1.6\%$ compressive strain over approximately three pseudocubic unit cells, yielding a strain gradient of $\sim 3.2 \times 10^7$~m$^{-1}$. Critically, the spatial extent of the tensile strain region tracks with the suppression of the Y/Y* Fourier amplitude, confirming that this strain feature is intrinsic to the APB itself rather than arising from the background strain field of the surrounding membrane. The hard wall spans approximately four atomic planes in the PLD mapping and Y/Y* Fourier amplitude, corresponding to a width of $\approx 1$-$1.5$~nm. Such a narrow wall implies large spatial gradients of the underlying rotational order parameter, scaling as $\nabla \phi \sim \phi_0/\xi$ where $\xi$ is the wall width. The freestanding geometry likely reduces long-range elastic constraints present in substrate-clamped films, permitting steeper interpolation of the tilt field and thereby enhancing these strain gradients. This sharp gradient directly drives the observed polar displacement via the flexoelectric constitutive relation,
\begin{equation}
  P_i = f_{ijkl}\,\frac{\partial \varepsilon_{jk}}{\partial x_l},
  \label{eq:flexo_polar}
\end{equation}
where $P_i$ is the induced polarisation, $\varepsilon_{jk}$ is the strain tensor, and $f_{ijkl}$ is the fourth-rank flexoelectric tensor \cite{zubkoFlexoelectricEffectSolids2013, tagantsevPiezoelectricityFlexoelectricityCrystalline1986}, consistent with the roto-flexoelectric coupling framework described in Supplemental Note~A.

At the easy (translation-equivalent) boundary, negligible strain is measured at the wall with only a mild tensile strain of $\sim 1\%$ present in the vicinity. The spatial distribution does not follow the wall and is not co-localised with any suppression of the Y/Y* Fourier amplitude. This background strain is attributable to the broader strain field associated with the embedded $90^\circ$ walls discussed in the following section, rather than to the easy boundary itself. Consistent with the Ising-like collapse of $\boldsymbol{\phi}$ at the easy wall, in which the simultaneous suppression of all tilt components prevents the $M_3^+$ component from amplifying the roto-flexoelectric coupling, no polar displacement above the experimental noise floor is associated with the easy boundary, in direct contrast to the hard wall response.

A key question raised by the observation of polar displacements in a metallic system is whether itinerant carrier screening is consistent with the spatial extent of the polar features. Metallic screening in SrRuO$_3$ operates on sub-unit-cell length scales: Drude-Lorentz fits to near-infrared optical spectroscopy yield a plasma frequency $\omega_p \approx 3.4$~eV and a background permittivity $\varepsilon_b \approx 2.67$ \cite{BraicSciRep2015}, placing the electrostatic screening length well below one unit cell dimension, consistent with the Fermi screening length of $\lambda \approx 0.5$~\AA{} used to model SrRuO$_3$ electrodes in ferroelectric device heterostructures \cite{Jiang2022}. Both estimates confirm that homogeneous electric dipoles are suppressed on the scale of a single lattice spacing. The polar displacements at hard antiphase boundaries extend over $\sim$8~nm, more than an order of magnitude beyond any electrostatically screened region. This is physically consistent because the polar displacements observed here are not driven by a static electric field but by a structural (flexoelectric) driving force whose spatial extent is set by the strain gradient field rather than by electrostatic screening. The strain gradient field at the hard wall extends over the same $\sim$8~nm length scale as the polar displacement, confirming that it is the mechanical driving force, not the electrostatic response, that defines the spatial extent of the polar texture.

\begin{figure}[hbt!]
\centering
\includegraphics[width=\columnwidth]{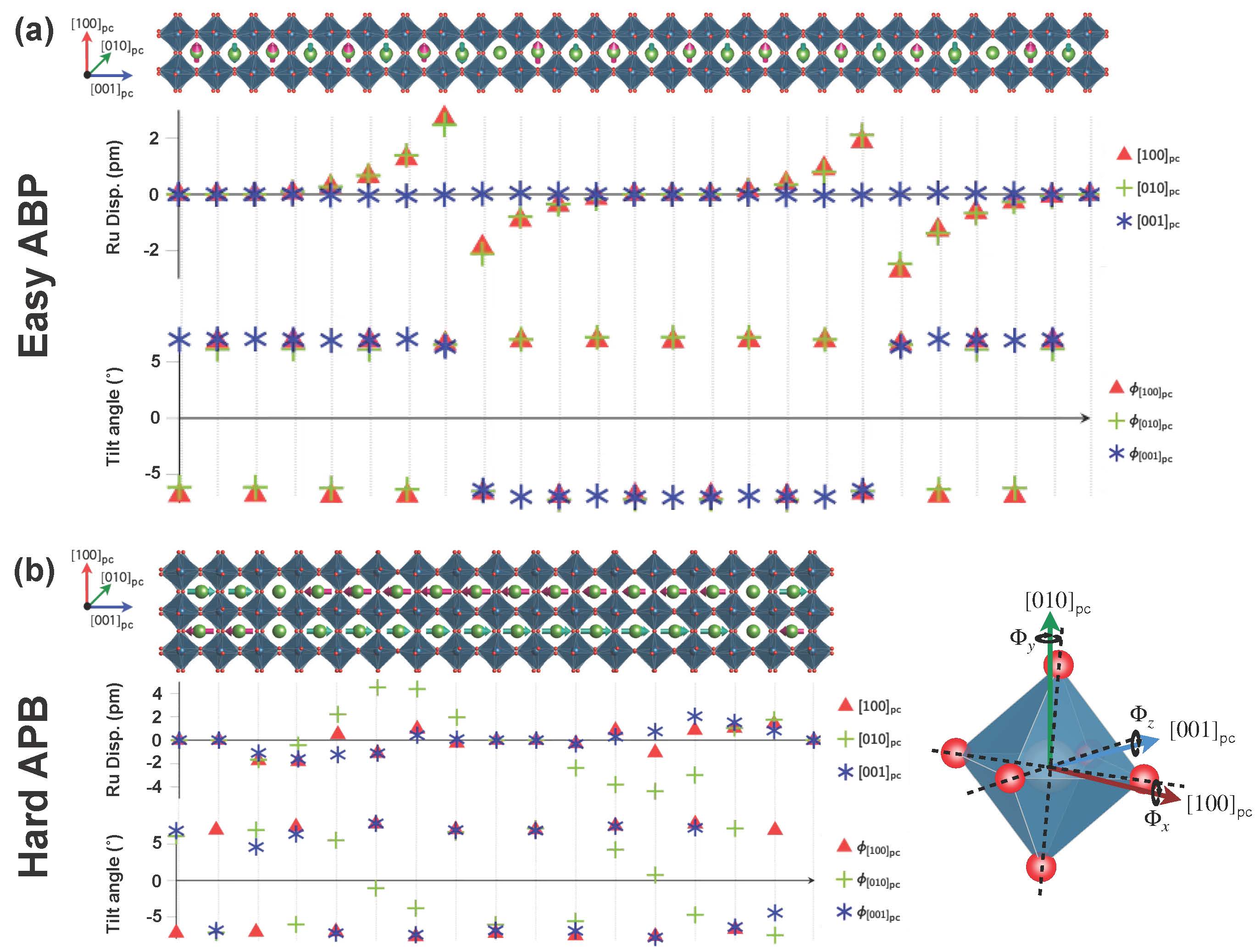}
\caption{\small
\textbf{First-principles structures and order-parameter profiles at easy and hard antiphase boundaries in SrRuO\textsubscript{3}.}
\textbf{(a)} DFT-relaxed supercell of the easy antiphase boundary. Upper panel: atomic structure with arrows indicating Sr displacements relative to their parent cubic positions, coloured by displacement direction. Lower panels: spatially resolved Ru off-centring displacement and octahedral tilt angle about the pseudocubic axes as a function of position across the boundary, extracted from the relaxed ionic coordinates. The Ising-like collapse of both quantities at the wall centre is consistent with the simultaneous suppression of all tilt components required by the discrete $\mathbf{t}_4$ translation.
\textbf{(b)} Equivalent DFT-relaxed supercell and profiles for the hard antiphase boundary. The Ru off-centring profiles reflect the N\'{e}el-like interpolation character of the wall, in which the antiphase tilt components pass through zero while the in-phase $M_3^+$ component is preserved, consistent with the c-flip domain relation.}
\label{fig: Theory Figure}
\end{figure}

First-principles calculations provide an atomistic test of the boundary-interpolation mechanism inferred from the STEM displacement maps of the boundaries. Supercells containing translation-equivalent easy APBs and translation-inequivalent hard APBs were relaxed within spin-polarised DFT+U, and the resulting ionic coordinates were used to extract Ru off-centring, strain, and octahedral tilt profiles across the walls (see full details in the SI) as show in Fig.~\ref{fig: Theory Figure}. For the easy boundary, the relaxed structure exhibits an Ising-like collapse: the Ru displacement and all tilt components are strongly suppressed at the wall centre, so the local dipole response remains weak and does not produce a robust polar boundary state. For the hard boundary, the relaxed structure instead shows a Néel-like interpolation, in which the antiphase tilt components reverse across the wall while the in-phase $M_{3}^{+}$ component is retained. This preserved rotational component coincides with enhanced Ru off-centring and a finite local dipole response, confirming that the hard boundary provides the structural conditions required for polarisation. The DFT results thus independently validate the experimental assignment of nonpolar easy APBs and polar hard APBs.

\subsection*{Roto-flexoelectric Coupling and Symmetry Origin of Polar Selectivity}

The symmetry origin of the selective polar response can be understood within the established framework for antiferrodistortive perovskites \cite{morozovskaInterfacialPolarizationPyroelectricity2012, schranzSignaturesPolarityFerroelastic2022a, weiFerroelectricTranslationalAntiphase2014}. As detailed above, the key quantity is the spatial gradient of the octahedral tilt order parameter $\boldsymbol{\phi} = (\phi_x, \phi_y, \phi_z)$, the primary order parameter to which the experimentally measured staggered Sr displacement $\mathbf{A}$ is linearly coupled through rotostriction. The Ising versus Néel interpolation character observed directly in $\mathbf{A}$ therefore provides experimental evidence for the behaviour of $\boldsymbol{\phi}$ across each boundary type, forming the experimental foundation for the symmetry argument that follows. Through rotostriction, spatial variation of $\boldsymbol{\phi}$ generates strain gradients, which in turn induce polarisation via the flexoelectric constitutive relation (Eq.~\ref{eq:flexo_polar}). Whether polarisation emerges at a given boundary is determined not simply by the presence of gradients, which exist at both boundary types, but by whether the multicomponent structure of the $a^-a^-c^+$ order parameter allows those gradients to drive a net polar response through the roto-flexoelectric coupling term $P_i \sim \lambda_{ijkl}\,\phi_j\,\partial_k \phi_l$.

At easy boundaries, the $180^\circ$ discontinuity in the staggered Sr modulation is accommodated by the discrete parent-cubic translation $\mathbf{t}_4 = \tfrac{1}{2}[110]_\mathrm{pc}$, which simultaneously reverses all three components of the tilt vector, both the antiphase $R_4^+$ components $(\phi_x, \phi_y)$ and the in-phase $M_3^+$ component $\phi_z$. The wall interpolation is therefore Ising-like: all components collapse toward zero together at the wall centre while the orientation of $\hat{\boldsymbol{\phi}}$ does not rotate, directly reflected in the experimentally observed Ising character of $\mathbf{A}$ at the easy boundary. Within the roto-flexoelectric coupling term, the gradients $\partial_k \phi_{x,y}$ and $\partial_k \phi_z$ are largest at the wall centre, but at this same point all components $\phi_j$ are simultaneously suppressed toward zero by the Ising collapse. The prefactor $\phi_j$ is therefore minimised precisely where $\partial_k \phi_l$ is maximised, strongly suppressing the net polar response. Experimentally, no polar displacement above the noise floor is detected at easy boundaries, consistent with this suppression.

At hard boundaries, the domain configuration is related by a c-flip operation that reverses $(\phi_x, \phi_y)$ while preserving $\phi_z$. No parent-cubic translation can simultaneously flip the $R_4^+$ components while leaving $M_3^+$ unchanged; the only translation producing a real-valued sign change, $\mathbf{t}_4$, reverses all three components and therefore cannot reconcile the hard-wall configuration. Continuous suppression of the antiphase tilt components across the wall is geometrically unavoidable, and the interpolation is N\'{e}el-like: $(\phi_x, \phi_y)$ pass through zero while $\phi_z$ is preserved and remains large throughout, directly reflected in the continuous rotation of $\hat{\mathbf{A}}$ and weak amplitude suppression of $|\mathbf{A}|$ observed experimentally at the hard boundary. The consequences for the roto-flexoelectric coupling are qualitatively different from the easy wall: the preserved $M_3^+$ component $\phi_z \approx \phi_z^0$ provides a large, non-zero prefactor precisely where the gradients $\partial_k \phi_{x,y}$ are steepest, amplifying rather than suppressing the coupling term and driving the strong interfacial polarisation observed experimentally. 

The Ising versus N\'{e}el interpolation character of the two boundary types is therefore not merely a phenomenological distinction but is mechanistically connected to the presence or absence of polarisation, arising directly from whether all tilt components reverse together or only the $R_4^+$ components pass through zero while $M_3^+$ is preserved. The properties of the two boundary classes are summarised in Table~\ref{tab:summary}; the full group-theoretical derivation, phase-factor analysis, and quantitative treatment of the coupling terms are provided in Supplemental Note~B. Polar translational boundaries have been seen in insulating antiferroelectric PbZrO\textsubscript{3} \cite{WeiPreferential, Wei2014, Rychetsky2019}. The present work establishes that analogous polar boundaries exist in a ferromagnetic metal. In addition, this work demonstrates that the Ising versus Néel interpolation character of antiphase boundaries, determined by the translational relationship between domains, constitutes the mechanistic origin of polar selectivity in multicomponent $a^-a^-c^+$ orthorhombic perovskites. The strain gradient evidence for each boundary class is shown directly in Fig.~\ref{fig: Long range Domain Mapping}d.

\begin{table}[htb!]
\centering
\begin{tabular}{lll}
\hline
Property & Easy boundary & Hard boundary \\
\hline
Domain relation        & Translation $\mathbf{t}_4 = \tfrac{1}{2}[110]_\mathrm{pc}$ 
                       & c-flip: no translation equivalent \\
$R_4^+$ sign change    & Yes (via $\mathbf{t}_4$) 
                       & Yes (via c-flip) \\
$M_3^+$ sign change    & Yes (via $\mathbf{t}_4$) 
                       & No (preserved by c-flip) \\
Wall interpolation     & Ising-like (all components $\to 0$) 
                       & N\'{e}el-like ($\phi_z$ preserved) \\
Order-param.\ gradient & Present, but $\phi_j \to 0$ at wall centre  
                       & Present, $\phi_z^0$ amplifies coupling \\
Roto-flexo coupling    & Suppressed 
                       & Amplified \\
Polarisation           & Absent 
                       & Present \\
\hline
\end{tabular}
\caption{Summary of easy and hard antiphase boundary properties in
$a^-a^-c^+$ SrRuO\textsubscript{3}.}
\label{tab:summary}
\end{table}

\subsection*{Polar Nanoclusters and Embedded 90$^\circ$ Walls}

\begin{figure}[hbt!]
\centering
\includegraphics[width=\columnwidth]{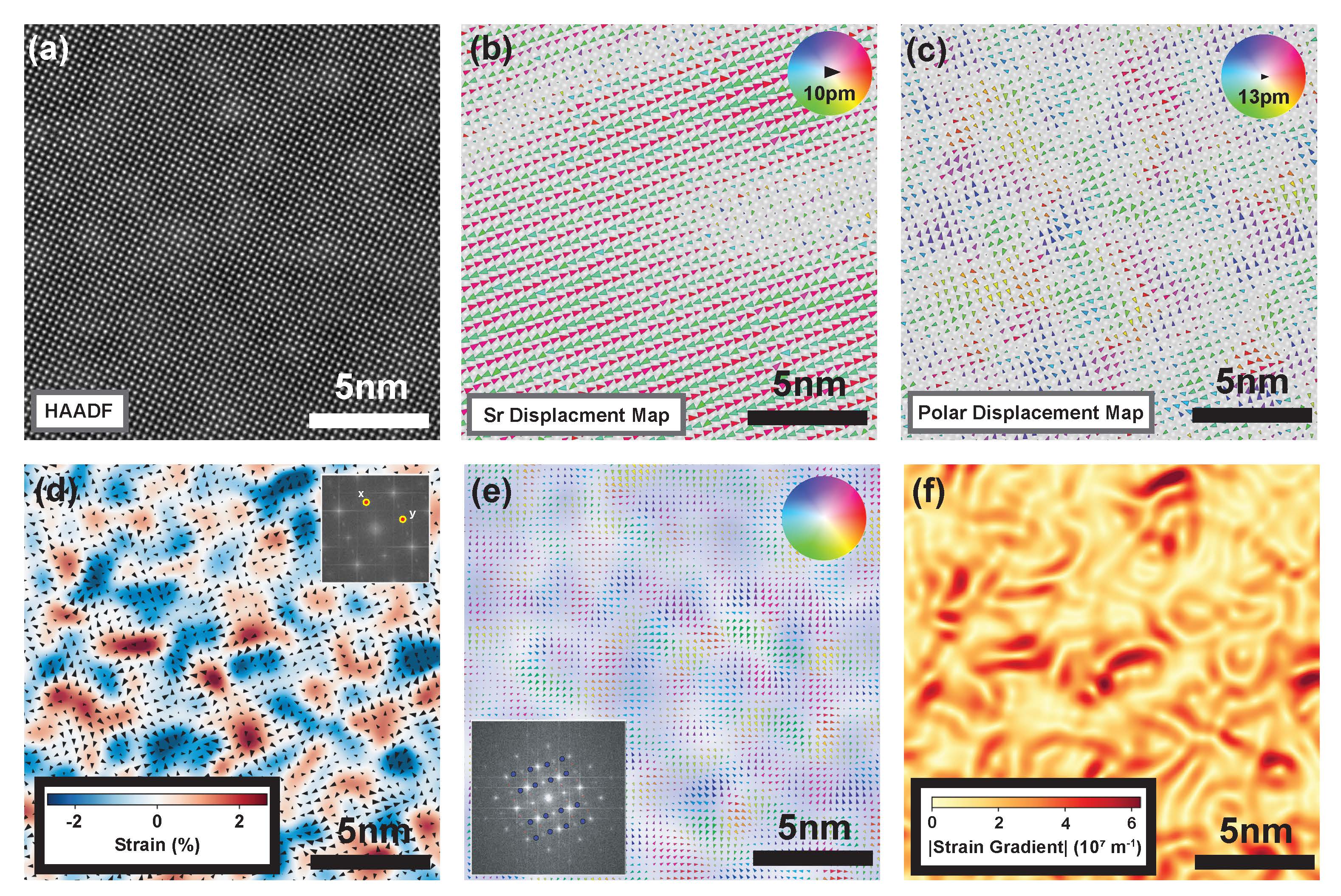}
\caption{\small 
\textbf{Polar nanoclusters and embedded 90$^\circ$ ferroelastic walls in a nominally single-domain region of freestanding SrRuO\textsubscript{3}.}
\textbf{(a)} ADF-STEM image of a region isolating a nominally single Y/Y$^*$ ferroelastic domain.
\textbf{(b)} Sr displacement map of the same region, revealing significant short-range structural complexity within the nominally uniform Y/Y$^*$ variant.
\textbf{(c)} Polar displacement map, defined by Ru displacement within the surrounding Sr cage, revealing a heterogeneous nanoscale polar texture composed of clustered in-plane polar regions distributed throughout the nominally single-domain region.
\textbf{(d)} Polar displacement vectors overlaid on the hydrostatic strain field. Polarisation vectors systematically orient along the negative hydrostatic strain gradient, flowing from regions of tensile toward compressive strain, consistent with flexoelectric coupling (Eq.~\ref{eq:flexo_polar}).
\textbf{(e)} X/X$^*$ Fourier amplitude overlaid with $-\nabla\varepsilon_{\mathrm{hyd}}$. Finite X/X$^*$ intensity within the nominal Y/Y$^*$ region demonstrates the presence of embedded 
$90^\circ$ ferroelastic domain walls. The spatial correlation between 
the X/X$^*$ amplitude boundaries and the peaks in $-\nabla\varepsilon_{\mathrm{hyd}}$ confirms co-localisation of both quantities at the ferroelastic wall.
\textbf{(f)} Magnitude of the hydrostatic strain gradient field.
}
\label{fig: Short range Polarisation}
\end{figure}

Fig.~\ref{fig: Short range Polarisation}a shows an ADF-STEM image of a region isolating a nominally single Y/Y* ferroelastic domain extracted from the long-range domain structure in Fig.~\ref{fig: Long range Domain Mapping}. The corresponding Sr displacement map (Fig.~\ref{fig: Short range Polarisation}b) reveals that although this region appears single-domain at long length scales, significant short-range structural complexity is evident within the nominally uniform variant. 

The polar displacement map (Fig.~\ref{fig: Short range Polarisation}c) reveals a heterogeneous nanoscale polar texture composed of clustered in-plane polar regions embedded within the metallic host. Fig.~\ref{fig: Autocorrelation} shows that autocorrelation of the hydrostatic strain map exhibits a pronounced peak at $\approx 4$~nm, defining a characteristic spatial periodicity of the strain modulation and its associated polar texture.

To understand the origin of this polar texture, we overlay the polar displacement map with the hydrostatic strain field (Fig.~\ref{fig: Short range Polarisation}d). A strong spatial correlation is immediately evident: polarisation vectors systematically orient along the negative hydrostatic strain gradient, flowing from regions of tensile strain toward compressive strain. The hydrostatic strain varies from approximately $+2\%$ to $-2\%$ over only a few nanometres, and the gradient of this strain field couples directly to polarisation through the flexoelectric constitutive relation (Eq.~\ref{eq:flexo_polar}), providing a physically direct driving term for the observed polar displacements and strongly supporting a flexoelectric origin of the nanoscale polar texture. The polar displacement field is therefore spatially heterogeneous in a manner that directly reflects the underlying strain field: nanoscale pockets of alternating tensile and compressive strain produce locally divergent and convergent polarisation vector configurations, with vector orientations tracking the sign and direction of the local strain gradient. The polarisation configurations observed around the strain pockets bear resemblance to the divergent and convergent configurations found within topological polar textures such as vortices and skyrmions reported in polar oxide systems.\cite{Das2019, Yadav2016}

The spatial heterogeneity of the polar texture thus reflects a competition between two length scales. Where strain gradients are large, the flexoelectric driving force dominates and polar displacements track the local strain gradient field directly, as demonstrated at the hard antiphase boundaries above. Where strain gradients are weaker, the flexoelectric driving force is reduced and longer-range dipole-dipole interactions become relatively more important, favouring more uniformly aligned polar vectors over larger domains. Superimposed on this competition is metallic screening, which sets a hard lower bound on the length scale over which coherent polar displacements can persist: any polar feature smaller than the sub-unit-cell electrostatic screening length of SrRuO$_3$ \cite{BraicSciRep2015} will be suppressed by itinerant carrier screening regardless of the local strain gradient. The observed polar texture is therefore the result of three competing interactions: flexoelectric coupling to the local strain gradient, longer-range dipole-dipole alignment, and metallic screening, whose relative strengths vary spatially across the ferroelastically twinned membrane and collectively determine the characteristic $\sim$4~nm length scale of the polar texture.

The structural origin of these large strain gradients is revealed in Fig.~\ref{fig: Short range Polarisation}e, which overlays the X/X$^*$ Fourier amplitude with $-\nabla\varepsilon_{\mathrm{hyd}}$. Finite X/X$^*$ intensity within the nominal Y/Y$^*$ region directly demonstrates the presence of embedded $90^\circ$ ferroelastic domain walls, structural boundaries at which the orthorhombic tilt axis rotates by $90^\circ$ within the plane of the membrane. At such a wall, the exchange of the two short orthorhombic axes $a_o$ and $c_o$ between in-plane directions produces a deviatoric eigenstrain mismatch whose elastic accommodation generates a localised hydrostatic strain gradient at the boundary; rotostriction as the tilt field interpolates continuously across the wall contributes an additional hydrostatic component through the same flexoelectric constitutive relation. The spatial correlation between the X/X$^*$ amplitude boundaries and the peaks in $-\nabla\varepsilon_{\mathrm{hyd}}$ confirms the co-localisation of both quantities at the ferroelastic wall itself, identifying the embedded $90^\circ$ walls as the source of the strain gradients driving the flexo-polar response. The magnitude of this strain-gradient field (Fig.~\ref{fig: Short range Polarisation}f) reaches values on the order of $10^7$~m$^{-1}$, comparable to that required for substantial flexoelectric polarisation in metallic SrRuO\textsubscript{3}, as demonstrated by Peng~\textit{et al.}\ \cite{RN415}, where strain gradients of similar magnitude induced measurable Ru off-centring and magnetoelectric modulation.

The structural assignment is further corroborated by site-specific Sr-Sr-Sr bond angle analysis and Fourier amplitude mapping of the same region (Fig.~\ref{fig: PLD_vs_Site_Analysis}), which yield domain boundary positions in quantitative agreement with those identified by Fourier filtering methods, demonstrating that real-space and reciprocal-space metrics report consistently on the same underlying structural heterogeneity. We further note that the polar displacement field exhibits spatial coherence directly tracking the strain gradient field over length scales of $\sim$4~nm, considerably larger than both the sub-unit-cell electrostatic screening length of SrRuO$_3$ \cite{BraicSciRep2015} and the projected unit cell dimension. Such coherent, strain-gradient-correlated polar structures are physically inconsistent with the random superposition of projection artefacts arising from uncorrelated domain overlap in the thickness direction, which would produce incoherent displacement noise rather than spatially organised polar textures. The robustness of the polar texture to local displacement noise is demonstrated by nearest-neighbour vector averaging in Fig.~\ref{fig: Smoothing}. 

The presence of nanoscale $90^\circ$ subvariants embedded within a nominally single ferroelastic Y/Y* domain may at first appear energetically costly. To complement the $90^\circ$ wall experimental findings in this work, DFT calculations were performed (Supporting Information Fig. \ref{fig: 90_Deg_DFT}) that reveal 5-6~pm displacement and thus agreement with the experimental STEM statistical analysis present in Fig. \ref{fig: Histogram}. It is also important to note that hierarchical twinning is a well-established strain-accommodation mechanism in ferroelastic systems, arising from the competition between domain-wall energy and long-range elastic compatibility~\cite{SaljeElasticInstabilities,SaljeFerroelasticTwinStructures,Khachaturyan1991}. Fine-scale twinning accommodates misfit at lower elastic energy cost than maintaining a homogeneous strain gradient, and is therefore energetically favoured in systems where long-range elastic strain would otherwise persist. 

In substrate-supported films, the rigid substrate introduces long-range elastic strain into the clamped lattice that suppresses fine-scale twin refinement. Upon membrane release, the free surfaces impose zero-traction boundary conditions, allowing this long-range elastic strain to relax. The residual driving force opposing fine-scale twin formation is consequently reduced, enabling the system to subdivide more readily into the nanoscale $90^\circ$ subvariants observed here. Additionally, the membrane geometry introduces curvature, as evidenced by the optical image of the film suspended over the TEM SiN hole (Fig.~\ref{fig: SRO_low_mag}), generating further spatially varying strain fields that promote hierarchical twinning as an accommodation mechanism. The observed hierarchical domain structure is thus energetically consistent with the microelastic framework of ferroelastic membranes and naturally accounts for the large strain gradients responsible for the emergent flexo-polar response.

\section*{Conclusion}

Polar order and metallicity are fundamentally at odds, yet we show that geometry alone can reconcile them. By releasing epitaxial SrRuO$_3$ from its substrate, ferroelastic domain refinement concentrates structural gradients at two classes of topological boundary, generating pervasive polar textures throughout a single freestanding conducting layer without twisted stacking, dedicated polar interfaces, or chemical symmetry breaking. The mechanism is quantitatively captured by a unified flexoelectric constitutive framework and independently confirmed by first-principles calculations.

The central mechanistic insight is that polar selectivity in multicomponent $a^-a^-c^+$ tilt systems is governed not by the presence of structural gradients, which exist at all boundary types, but by the interpolation character of the order parameter across the wall. At translation-inequivalent boundaries, N\'{e}el-like interpolation preserves a finite $M_3^+$ prefactor precisely where tilt gradients are steepest, amplifying roto-flexoelectric coupling and driving strong interfacial polarisation. At translation-equivalent boundaries, Ising-like collapse suppresses all tilt components simultaneously, extinguishing the polar response. This distinction, determined entirely by the translational relationship between adjacent domains, constitutes a symmetry-based design rule: the Ising or N\'{e}el character of a domain wall is not a material property but a boundary property, and is therefore addressable through domain engineering without changing the host material.

These findings identify ferroelastic domain topology, and specifically the interpolation character of multicomponent order parameters, as the key factor governing polarisation in this system. More broadly, they establish a general route to polar metals in which intrinsic structural gradients stabilised by freestanding geometries and translation inequivalent domain pairing replace the need for interfaces or chemical asymmetry. This provides a scalable materials design principle for engineering reconfigurable polar and spin functional behaviour across the perovskite family

\printbibliography

\section*{Methods}

\subsection*{Epitaxial film growth}

Epitaxial heterostructures were deposited by pulsed laser deposition using a Neocera Pioneer 120 system equipped with a 248\,nm KrF excimer laser. The base pressure prior to growth was $1 \times 10^{-6}$\,Torr. Commercial TiO$_2$-terminated (001)-oriented SrTiO\textsubscript{3} substrates (Shinkosha) were used as received.

A Sr\textsubscript{3}Al$_2$O$_6$ (SAO) sacrificial layer of thickness 18\,nm, determined by cross-sectional STEM, was grown at 920$^\circ$C under 1.2\,mTorr O$_2$ at 3\,Hz with a laser fluence of 1.3\,J\,cm$^{-2}$. Subsequently, SrRuO\textsubscript{3} (SRO) was deposited at 680$^\circ$C under 65\,mTorr O$_2$ at 3\,Hz with a fluence of 1.0\,J\,cm$^{-2}$. The SRO thickness was 34\,nm, determined by X-ray reflectivity (XRR). After deposition, samples were cooled to room temperature at 10$^\circ$C\,min$^{-1}$ under 65\,mTorr O$_2$.

\subsection*{X-ray diffraction and reflectivity}

High-resolution X-ray diffraction (XRD) and X-ray reflectivity (XRR) were performed using a Malvern Panalytical Empyrean MultiCore diffractometer equipped with a Cu K$_\alpha$ source (40\,kV, 40\,mA) and a Ge crystal monochromator with parallel-beam optics. Structural and morphological characterisation data are shown in Fig.~\ref{fig: All_XRD_AFM}.

Specular XRR measurements yielded an SRO thickness of 34\,nm. Reciprocal space maps confirmed epitaxial alignment and were used to assess the strain state prior to membrane release.

\subsection*{Atomic force microscopy}

Surface morphology was characterised in tapping mode using an Oxford Instruments Asylum Research MFP-3D Origin™ atomic force microscope with NuNano SCOUT-70 Si probes. Images were analysed to quantify root mean squared surface roughness over a representative $5 \times 5\,\mu\text{m}^2$ area.

\subsection*{Freestanding membrane fabrication}

Freestanding SRO membranes were obtained by selective dissolution of the water-soluble SAO sacrificial layer. A 300\,$\mu$m-thick polydimethylsiloxane (PDMS) support layer was laminated onto the as-grown heterostructure to provide mechanical stability during release. Samples were immersed in deionised water for 3\,hr 5to dissolve the SAO layer, resulting in delamination of the SRO film. Upon release, elastic strain relaxation induced controlled cracking, yielding isolated membrane flakes supported by PDMS.

The released flakes were transferred onto DENSsolutions Lightning Nano-Chips. The MEMS chip was heated to 80$^\circ$C to promote adhesion, and the PDMS was slowly peeled away, leaving the SRO membrane suspended across the chip window.

\subsection*{Electron channelling contrast imaging}
Scanning electron microscopy was performed on a TESCAN AMBER-X instrument, which is equipped with a Schottky Field Emission Gun, a 4 quadrant backscatter dlectron detector, and a double tilt `rocking' stage. Electron channelling contrast was obtained using the `channelling' mode to collect an electron channelling pattern (ECP). For the `clamped' film, the sample was tilted to place the centre of the [001] zone axis along the optical axis of the electron microscope. For the freestanding film, the ECCI was performed with the [001] Si zone axis of the wafer aligned along the optical axis (as the freestanding film was made of regions too small to collect a single selected area electron channelling pattern from). To ensure that the contrast obtained was reproducible for the freestanding film, multiple other imaging conditions were tested to verify ECCI contrast.

\subsection*{Scanning transmission electron microscopy}

Aberration-corrected STEM was performed using a Thermo Fisher Scientific Spectra 300 operated at 300\,kV. ADF-STEM images were acquired with a convergence semi-angle of 20\,mrad and a collection angle range of 39-200\,mrad. Polar displacements were extracted exclusively from Sr and Ru cation column positions in ADF micrographs. This approach is robust against the misorientation-induced artefacts that affect oxygen 
column positions in annular bright-field imaging, where specimen tilts of only a few milliradians can introduce artificial displacements exceeding 10~pm~\cite{LiABF2017}. The beam current was ~42\,pA with a dwell time of 125\,ns per pixel. Electron dose was carefully controlled to minimise beam-induced artefacts. Sequential image stacks were aligned using rigid registration as described by Savitzky \textit{et al.}\cite{RN474} prior to quantitative lattice analysis.

\subsection*{Strain mapping and phase analysis}

Quantitative lattice displacement and strain mapping were performed on registered ADF-STEM datasets using the KEMSTEM framework \cite{schnitzerQuantitativeApproachesMultiscale2025}. Atomic column positions were extracted through two-dimensional peak fitting, and local lattice vectors were computed to generate spatially resolved strain tensors relative to an internal reference region within each membrane. This internal referencing minimises residual global distortions arising from membrane curvature or scan artefacts.

To disentangle coexisting structural distortions, phase lock-in analysis was implemented following Goodge \textit{et al.}\cite{goodgeDisentanglingCoexistingStructural2022a}. Reciprocal lattice vectors were isolated in Fourier space and demodulated to reconstruct real-space displacement fields with sub-picometre sensitivity. Periodic lattice displacement fields associated with embedded ferroelastic domain walls were further analysed using phase-sensitive techniques developed by Savitzky \textit{et al.}\cite{RN474}. This approach enables quantitative mapping of symmetry-breaking distortions and nanoscale strain modulations across 90$^\circ$ domain walls. Strain precision was estimated to be 0.1\%, consistent with previously reported performance of phase lock-in STEM analysis.

\subsection*{Theory calculations}
To complement the experimental findings in this work, a theoretical study of antiphase boundaries in SrRuO$_{3}$ was undertaken with the framework of Kohn-Sham density functional theory (DFT)~\cite{Kohn1965} using the Vienna \textit{ab initio} Simulation Package code (VASP 6.5.1).~\cite{Kresse1996a,Kresse1996b} This approach employs periodic boundary conditions with a plane-wave basis set to represent the electronic structures of extended materials, using a set of projector augmented wave (PAW) potentials with valence electrons as follows: Sr (4\textit{s}$^2$\,4\textit{p}$^6$\,4\textit{d}$^{0.001}$\,5\textit{s}$^{1.999}$), Ru (4\textit{p}$^6$\,4\textit{d}$^7$\,5\textit{s}$^1$) and O (2\textit{s}$^2$\,2\textit{p}$^4$).~\cite{Blochl1994,Kresse1999} The effects of electronic exchange and correlation were accounted for using the PBEsol density functional approximation~\cite{Perdew2008} - a revised form of the Perdew-Burke-Ernzerhof~\cite{Perdew1996,Perdew1997} generalised gradient approximation (GGA) which yields improved equilibrium properties for solid-state chemical systems. In this study, the spin unrestricted formalism of DFT was used and the Gaussian smearing method was applied to the occupation numbers of electronic states around the Fermi energy with a width of $0.01$ eV. In addition, to account for the strong correlation arising due to interactions within the 4\textit{d} electrons localised on the Ru ions, the DFT+U approach of Dudarev is employed with an effective on-site Coulomb interaction parameter $U_\mathrm{eff} = U - J = 1$ eV for Ru ions.~\cite{Dudarev1998} 
	
	For all calculations, a plane-wave cut-off energy of $700$ eV was employed whilst the Brilloun zone was integrated using a $\Gamma$-centered reciprocal space grid with a density of 0.228 \AA$^{-1}$; these were chosen yield convergence in the energy of the SrRuO$_{3}$ unit cell in the orthorhombic phase (space group \textit{Pbnm}) to within $1$ meV per atom. Initially, the equilibrium structure of orthorhombic SrRuO$_{3}$ was obtained through relaxation of ionic positions and lattice parameters using a conjugate gradient algorithm, reaching convergence only when the residual force acting on each ion decreased below $0.1$ meV \AA$^{-1}$. This resulted in an orthorhombic unit cell with lattice parameters of $a=5.510$\AA{}, $b=5.546$\AA{} and $c=7.829$\AA{}; this is in close agreement with experimental measurements obtained at $1.5$K of $a=5.532$\AA{}, $b=5.566$\AA{} and $c=7.845$\AA{}.~\cite{Bushmeleva2006}
	
	To construct a model of the easy antiphase boundary, a supercell of dimension $1\times1\times12$ was constructed from the relaxed SrRuO$_{3}$ unit cell, resulting in a structure effectively comprising of $1\times1\times24$ pseudocubic units. The antiphase structure was constructed by applying a transformation to the original unit cell described in Eq.~\eqref{eq:easy_ap_trans}.  
	\begin{equation}\label{eq:easy_ap_trans}
		\begin{pmatrix}
			\mathbf{a}^{\prime} \\
			\mathbf{b}^{\prime} \\
			\mathbf{c}^{\prime} \\
			1
		\end{pmatrix} = 
		\begin{pmatrix}
			1 & 0 & 0 & \frac{1}{2} \\
			0 & 1 & 0 & \frac{1}{2} \\
			0 & 0 & -1 & 0 \\
			0 & 0 & 0 & 1
		\end{pmatrix}
		\begin{pmatrix}
			\mathbf{a} \\
			\mathbf{b} \\
			\mathbf{c} \\
			1
		\end{pmatrix}
	\end{equation}
	A supercell of dimension $1\times1\times6$ was constructed from the resulting antiphase unit cell and was substituted into the larger $1\times1\times12$ supercell, replacing the structure between the original unit cells with indices $(1,1,3\frac{1}{2})$ and $(1,1,9\frac{1}{2})$ (i.e. replacing pseudocubic units $8$ to $19$), yielding two easy antiphase boundaries, necessary to maintain periodicity along the $\mathbf{c}$ axis. The resulting structure was subsequently relaxed using the same methodology as for the unit cell, but with the atoms in the unit cells with indices $(1,1,1)$ and $(1,1,7)$ frozen in the two different phase configurations of the material, and with a convergence threshold of $0.01$ eV \AA$^{-1}$ for the residual force acting on each ion. 
	
	A similar approach was followed for constructing a model of the hard antiphase boundary. However, the orthorhombic unit cell was first transformed into the pseudocubic coordinate axis by the coordinate transformation 
	\begin{equation}\label{eq:hard_pc_trans}
		\begin{pmatrix}
			\mathbf{a}^{\prime} \\
			\mathbf{b}^{\prime} \\
			\mathbf{c}^{\prime} \\
			1
		\end{pmatrix} = 
		\begin{pmatrix}
			1 & 1 & 0 & \frac{1}{2} \\
			-1 & 1 & 0 & 0 \\
			0 & 0 & 1 & 0 \\
			0 & 0 & 0 & 1
		\end{pmatrix}
		\begin{pmatrix}
			\mathbf{a} \\
			\mathbf{b} \\
			\mathbf{c} \\
			1
		\end{pmatrix}
	\end{equation}
	from which the antiphase structure is obtained by applying a further coordinate shift of $\frac{1}{2}\mathbf{c}$. An $8\times1\times1$ supercell of the pseudocubic-transformed unit cell was constructed and the atoms in unit cells with indices $(2\frac{1}{2},1,1)$ to $(6,1,0)$ substituted with the equivalent size supercell of the antiphase structure. Atoms in the first and 9$^{th}$ pseudocubic units along the supercell were frozen and the structure relaxed in the same was as for the easy antiphase boundary. Finally, the 90$^\circ$ antiphase boundary was constructed in a near-identical manner, with the antiphase arrangement constructed by rotating the pseudocubic-transformed unit cell by 90$^\circ$ ($\mathbf{a}\to\mathbf{-c}$, $\mathbf{c}\to\mathbf{-a}$) rather than by a coordinate shift as for the hard antiphase boundary. 

\subsection*{Data availability}

The data that support the findings of this study are available \href{https://doi.org/10.5281/zenodo.19862672}{https://doi.org/10.5281/zenodo.19862672}.

\subsection*{Code availability}

Image registration was performed using the rigid registration implementation described by Savitzky \textit{et al.} \cite{savitzkyImageRegistrationLow2018} which can be found at \href{https://github.com/kem-group/rigidRegistration}{https://github.com/kem-group/rigidRegistration}. Strain mapping and phase lock-in analysis were conducted using the open-source KEMSTEM package \cite{schnitzerQuantitativeApproachesMultiscale2025} which can be found at \href{https://github.com/noahschnitzer/kemstem}{https://github.com/noahschnitzer/kemstem}. Additional custom analysis scripts are available from the corresponding author upon reasonable request.

\section{Author contributions}
R.H., Y.L., S.L.P, and A.H. grew the thin films. T.B.B. carried out the SEM ECCI experiments. M.S.C. carried out the STEM experiments. R.H., N.S., and G.T. analysed the STEM data. T.J.P.I. carried out the DFT calculations, E.B., S.M.G. and K.I. supported the theory work. M.S.C. led the project. R.H. and M.S.C. wrote the manuscript, with input and review from all authors.

\section*{Acknowledgements}
This work was made possible by the EPSRC Cryo-Enabled Multi-microscopy for Nanoscale Analysis in the Engineering and Physical Sciences EP/V007661/1. The AMBER-X pFIB-SEM hosted within the Electron Microscopy Lab at UBC, Vancouver Campus, was funded by the BCKDF and CFI-IF (\#39798: AM+). T.B.B. acknowledges funding from the Natural Sciences and Engineering Research Council of Canada (NSERC) [Discovery grant: RGPIN2022-04762, ‘Advances in Data Driven Quantitative Materials Characterization’]. R.H., G.T. and S.L.P acknowledge funding from the EPSRC Centre for Doctoral Training in the Advanced Characterisation of Materials (CDTACM)(EP/S023259/1) and R.H. and G.T. thank Cameca Ltd. for co-funding their PhD. N.S. and M.S.C. acknowledge funding from the ERC CoG DISCO grant 101171966. M.P., Y.L. and M.S.C. acknowledge funding from the Royal Society Tata University Research Fellowship (URF\textbackslash R1\textbackslash 201318) and Royal Society Enhancement Award RF\textbackslash ERE\textbackslash210200EM1.
This work was supported by the U.S. Department of Energy, Office of Science, Office of Basic Energy Sciences,
Materials Sciences and Engineering Division under Contract No. DE-AC02-05CH11231 within the Theory of Materials program. Computational resources were provided by the National Energy Research Scientific Computing Center and the Molecular Foundry, DOE Office of Science User Facilities supported by the Office of Science, U.S.
Department of Energy under Contract No. DE-AC02-05CH11231 and Sigma2 – the National Infrastructure for High-Performance Computing and Data Storage in Norway through project NN9264K. The work performed at the Molecular Foundry was supported by the Office of Science, Office of Basic Energy Sciences, of the U.S. Department of Energy under the same contract.

\newpage

\newrefsection

\begin{center}

\large \textbf{Supporting information}
\vspace{1em}

\large \textbf{Polar Topologies in a Ferroelastic Metal Membrane}

\vspace{0.5em}

Rahil Haria,$^{1}$ 
Noah Schnitzer,$^{1}$ 
T.\ Ben Britton,$^{1,2}$ 
Yaqi Li,$^{1}$ 
Tom J. P. Irons,$^{3}$ 
Sophia Linssen-Pitsaros,$^{1,4,5}$ 
Ella Banyas,$^{6,7*}$ 
Geri Topore,$^{1}$ 
Annabel Hoyes,$^{1}$ 
Mariana Palos,$^{1}$ 
Sin\'{e}ad M.\ Griffin,$^{6,7*}$ 
Katherine Inzani,$^{3*}$ 
Michele Shelly Conroy$^{1*}$

\vspace{0.5em}

{\small
$^{1}$Department of Materials, Imperial College London, London, SW7 2AZ, 
United Kingdom\\
$^{2}$Department of Materials Engineering, UBC, Vancouver, 
British Columbia, Canada\\
$^{3}$School of Chemistry, University of Nottingham, 
Nottingham NG7 2RD, UK\\
$^{4}$Department of Physics and Astronomy, University College London, 
Gower Street, London WC1E 6BT, United Kingdom\\
$^{5}$London Centre for Nanotechnology, 17-19 Gordon Street, 
London WC1H 0AH, United Kingdom\\
$^{6}$Materials Sciences Division, Lawrence Berkeley National Laboratory, 
Berkeley, CA 94720, USA\\
$^{7}$Molecular Foundry, Lawrence Berkeley National Laboratory, 
Berkeley, CA 94720, USA\\
}

\vspace{0.3em}

{\small *mconroy@imperial.ac.uk, katherine.inzani1@nottingham.ac.uk, 
sgriffin@lbl.gov}
\end{center}

\vspace{1em}

\renewcommand{\thefigure}{S\arabic{figure}}
\setcounter{figure}{0}
\renewcommand{\thetable}{S\arabic{table}}
\setcounter{table}{0}
\renewcommand{\figurename}{Figure}

\subsection*{Supplemental Note A: Roto-flexoelectric Coupling}

Orthorhombic SrRuO\textsubscript{3} adopts the \textit{Pbnm} structure
(space group No.~62, axis setting $c > a \approx b$; this is the same
space group as \textit{Pbnm} with the axes permuted as
$\mathbf{a}_\text{Pbnm} \leftrightarrow \mathbf{c}_\text{Pbnm}$).
The structure is characterised by a two-component antiferrodistortive
(AFD) order parameter
\begin{equation}
 \boldsymbol{\phi} = (\phi_x,\,\phi_y,\,\phi_z),
\end{equation}
where $(\phi_x, \phi_y)$ describe the \textit{antiphase} RuO\textsubscript{6}
octahedral rotations (irrep $R_4^+$ at the cubic Brillouin-zone corner
$\mathbf{k}_R = \tfrac{1}{2}[111]_\text{pc}$) and $\phi_z$ describes
the \textit{in-phase} rotation (irrep $M_3^+$ at the zone-edge point
$\mathbf{k}_M = \tfrac{1}{2}[110]_\text{pc}$).
All components are referred to pseudocubic axes
$[\hat{x},\hat{y},\hat{z}]_\text{pc}$.
Although the bulk structure is centrosymmetric, spatial variations of
$\boldsymbol{\phi}$ at ferroelastic domain walls and antiphase boundaries
reduce the local symmetry and can induce polarisation through
gradient-mediated coupling
\cite{zubkoFlexoelectricEffectSolids2013,tagantsevPiezoelectricityFlexoelectricityCrystalline1986}.

\subsubsection*{Free energy}

Within a Landau-Ginzburg-Devonshire formalism, the free energy
density is written
\begin{equation}
 F = F_{\phi} + F_P + F_\text{el} + F_\text{coup},
\end{equation}
where the structural (AFD) contribution is
\begin{equation}
 F_{\phi} = \frac{\alpha}{2}\phi_i\phi_i
    + \frac{\beta}{4}(\phi_i\phi_i)^2
    + \frac{g}{2}\frac{\partial\phi_i}{\partial x_j}
    \frac{\partial\phi_i}{\partial x_j},
\end{equation}
with $\alpha$ and $\beta$ the Landau coefficients and $g$ the gradient
stiffness governing domain-wall width.
The polar contribution is
\begin{equation}
 F_P = \frac{a}{2}P_iP_i + \frac{b}{4}(P_iP_i)^2,
\end{equation}
and the elastic energy is
\begin{equation}
 F_\text{el} = \tfrac{1}{2}\,c_{ijkl}\,u_{ij}\,u_{kl},
\end{equation}
where $u_{ij}$ is the symmetric strain tensor and $c_{ijkl}$ the
elastic stiffness tensor.

\subsubsection*{Coupling terms}

Two coupling terms are central to interfacial polarisation.
\textit{Rotostriction} couples octahedral tilts quadratically to strain,
\begin{equation}
 F_\text{roto} = -\lambda_{ijkl}\,u_{ij}\,\phi_k\,\phi_l,
 \label{eq:rotostriction}
\end{equation}
where $\lambda_{ijkl}$ is the rotostrictive tensor
\cite{morozovskaInterfacialPolarizationPyroelectricity2012}.
Because $\phi_i\phi_j$ varies spatially across a domain wall, spatial
variation of $\boldsymbol{\phi}$ necessarily generates strain gradients
through Eq.~\eqref{eq:rotostriction}.
\textit{Flexoelectric coupling} then links those strain gradients
linearly to polarisation,
\begin{equation}
 F_\text{flexo} = -f_{ijkl}\,P_i\,\frac{\partial u_{jk}}{\partial x_l},
\end{equation}
where $f_{ijkl}$ is the flexoelectric tensor
\cite{tagantsevPiezoelectricityFlexoelectricityCrystalline1986,%
zubkoFlexoelectricEffectSolids2013}.
Minimising the total free energy with respect to $P_i$ yields an
induced polarisation proportional to gradients of the structural order
parameter,
\begin{equation}
P_i \sim f_{ijkl}\,\lambda_{jkmn}\,\phi_m\,\frac{\partial\phi_n}{\partial x_l},
\label{eq:roto-flexo}
\end{equation}
demonstrating that polarisation emerges wherever the octahedral tilt field $\boldsymbol{\phi}$ varies spatially.

\newpage
\subsection*{Supplemental Note B: Octahedral-tilt transformation and 
translational compatibility at antiphase boundaries}

\subsubsection*{Order parameter and its transformation law}

In orthorhombic \textit{Pbnm} SrRuO\textsubscript{3} (Glazer $a^-a^-c^+$) the octahedral tilt order parameter introduced in the main text and Supplemental Note~A as $\boldsymbol{\phi} = (\phi_x,\phi_y,\phi_z)$ is here denoted $\mathbf{Q} = (Q_x, Q_y, Q_z)$ to distinguish the field itself from the transformation operations acting upon it; the two symbols refer to the same physical quantity throughout. $(Q_x, Q_y)$ are the antiphase $R_4^+$ components and $Q_z$ is the in-phase $M_3^+$ component, all referred to pseudocubic axes. Because octahedral rotations are axial vectors, they transform under a symmetry operation $R$ (with $\det R = \pm 1$) as
\begin{equation}
 \mathbf{Q}' = \det(R)\,R\,\mathbf{Q}.
 \label{eq:axial}
\end{equation}
The $\det(R)$ prefactor is essential: it ensures that an improper operation (mirror or inversion, $\det R = -1$) introduces an additional global sign flip relative to a polar vector. The staggered A-site displacement $\mathbf{A}$ introduced in the main text is a distinct secondary order parameter coupled to $\mathbf{Q}$ through rotostriction, with $\mathbf{A} \propto \boldsymbol{\phi}$ to leading order; it serves as the experimentally accessible proxy for $\mathbf{Q}$ but is not the primary tilt order parameter.

\subsubsection*{The c-flip operation}

We consider the mirror operation that reverses only the pseudocubic
$\hat{z}$ direction,
\begin{equation}
 R_z =
 \begin{pmatrix} 1 & 0 & 0 \\ 0 & 1 & 0 \\ 0 & 0 & -1 \end{pmatrix},
 \qquad \det(R_z) = -1.
\end{equation}
Applying Eq.~\eqref{eq:axial},
\begin{equation}
 \mathbf{Q}' = \det(R_z)\,R_z\,\mathbf{Q}
    = (-1)
     \begin{pmatrix}Q_x\\Q_y\\-Q_z\end{pmatrix}
     = (-Q_x,\,-Q_y,\,+Q_z).
  \label{eq:cflip}
\end{equation}
This operation reverses the two antiphase $R_4^+$ components while
\textit{preserving} the in-phase $M_3^+$ component.
The two domain types observed at the antiphase boundaries therefore
differ in the sign of $(Q_x,Q_y)$ but share the same $Q_z$.

\subsubsection*{Action of parent-cubic translations on the order parameter}

When a real-space translation $\mathbf{t}$ is applied, each modulated
order parameter component $Q$ at wavevector $\mathbf{k}$ acquires a
phase factor $e^{2\pi i\,\mathbf{k}\cdot\mathbf{t}}$ (coordinates in
units of $a_\text{pc}$).
For the factor to equal $-1$ (a sign flip) we need
$\mathbf{k}\cdot\mathbf{t} = \tfrac{1}{2}$.
Table~\ref{tab:phases} lists the phase factors for all seven distinct
primitive half-unit-cell translations of the parent cubic lattice.

\begin{table}[htbp]
\centering
\begin{tabular}{ccccc}
\hline
Label & $\mathbf{t}$ (pc fractional) &
 $e^{2\pi i\,\mathbf{k}_R\cdot\mathbf{t}}$ &
 $e^{2\pi i\,\mathbf{k}_M\cdot\mathbf{t}}$ &
 Effect on $(Q_x,Q_y,Q_z)$ \\
\hline
$\mathbf{t}_1$ & $(\tfrac{1}{2},0,0)$       & $i$ & $i$ & not a real-valued map \\
$\mathbf{t}_2$ & $(0,\tfrac{1}{2},0)$       & $i$ & $i$ & not a real-valued map \\
$\mathbf{t}_3$ & $(0,0,\tfrac{1}{2})$       & $i$ & $+1$ & not a real-valued map \\
$\mathbf{t}_4$ & $(\tfrac{1}{2},\tfrac{1}{2},0)$ & $-1$ & $-1$ & $(-Q_x,-Q_y,-Q_z)\to(+Q_x,+Q_y,+Q_z)$ \\
$\mathbf{t}_5$ & $(\tfrac{1}{2},0,\tfrac{1}{2})$ & $-1$ & $i$ & not a real-valued map \\
$\mathbf{t}_6$ & $(0,\tfrac{1}{2},\tfrac{1}{2})$ & $-1$ & $i$ & not a real-valued map \\
$\mathbf{t}_7$ & $(\tfrac{1}{2},\tfrac{1}{2},\tfrac{1}{2})$ & $-i$ & $-1$ & not a real-valued map \\
\hline
\end{tabular}
\caption{Phase factors acquired by each order-parameter component
under the seven non-trivial half-unit-cell translations of the parent
cubic lattice. $\mathbf{k}_R = \tfrac{1}{2}[111]_\text{pc}$ (antiphase,
$R_4^+$); $\mathbf{k}_M = \tfrac{1}{2}[110]_\text{pc}$ (in-phase,
$M_3^+$). A factor of $-1$ corresponds to a sign flip of that component;
factors of $\pm i$ are not real sign flips and cannot map a real order
parameter to itself. \label{tab:phases}}
\end{table}

The table reveals a key result: the \textit{only} half-unit-cell
translation that produces a real-valued sign change of any order
parameter component is $\mathbf{t}_4 = \tfrac{1}{2}[110]_\text{pc}$,
and it flips \textit{both} $R_4^+$ components \textit{and} the $M_3^+$
component simultaneously.

\subsubsection*{Easy boundary}

For the easy boundary, the two domains are related by the discrete translation $\mathbf{t}_4 = \tfrac{1}{2}[110]_\text{pc}$. As shown in Table~\ref{tab:phases}, this translation reverses all three order-parameter components:
\begin{equation}
  (-Q_x,\,-Q_y,\,-Q_z)
  \xrightarrow{\;\mathbf{t}_4\;}
  (+Q_x,\,+Q_y,\,+Q_z).
  \label{eq:easy}
\end{equation}
Crucially, \textit{both} the antiphase $R_4^+$ components \textit{and} the in-phase $M_3^+$ component reverse across the easy boundary. The experimentally observed reversal of the staggered Sr ($X_5^+$ mode) displacement pattern is consistent with this result, since both $R_4^+$
and $M_3^+$ distortions contribute to the staggered Sr displacement; the combined sign flip reproduces the observed phase reversal.

Because all three components reverse together, the wall interpolation is Ising-like: the magnitude $|\mathbf{Q}|$ collapses toward zero at the wall centre while the orientation $\hat{\mathbf{Q}}$ does not rotate, consistent with the experimental observation of strong amplitude suppression and negligible directional rotation of the staggered Sr order parameter at the easy boundary
(Fig.~\ref{fig: Long range Domain Mapping}c). This Ising collapse has direct consequences for the roto-flexoelectric coupling term,

\begin{equation}
  P_i \sim \lambda_{ijkl}\,Q_j\,\partial_k Q_l.
  \label{eq:roto-flexo-easy}
\end{equation}
The gradients $\partial_k Q_{x,y}$ and $\partial_k Q_z$ are largest at the wall centre, where the interpolation is steepest. However, at this same point all components $Q_j$ are simultaneously suppressed toward zero by the Ising collapse. The prefactor $Q_j$ in the coupling term is therefore minimised precisely where $\partial_k Q_l$ is maximised, strongly suppressing the net polar response. Experimentally, no polar displacement above the noise floor is detected at easy boundaries, suggesting that any residual polar response lies below the $\sim$2~pm detection limit of the present STEM analysis.

\subsubsection*{Hard boundary}

For the hard boundary, the two domains are related by the c-flip operation (Eq.~\eqref{eq:cflip}), which gives $(Q_x,Q_y,Q_z) \to (-Q_x,-Q_y,+Q_z)$. We now ask whether any primitive translation of the parent cubic lattice can map $(-Q_x,-Q_y,+Q_z) \to (+Q_x,+Q_y,+Q_z)$, i.e.,
whether it can flip the sign of the $R_4^+$ components while leaving $Q_z$ unchanged.

This requires simultaneously:
\begin{align}
  e^{2\pi i\,\mathbf{k}_R\cdot\mathbf{t}} &= -1
   \quad\text{(flip }R_4^+\text{)}, \\
  e^{2\pi i\,\mathbf{k}_M\cdot\mathbf{t}} &= +1
   \quad\text{(preserve }M_3^+\text{)}.
\end{align}
Inspection of Table~\ref{tab:phases} shows that \textit{no} primitive half-unit-cell translation satisfies both conditions simultaneously. The only translation yielding $-1$ for $\mathbf{k}_R$ with a real-valued result is $\mathbf{t}_4$, which also yields $-1$ for $\mathbf{k}_M$
and would therefore flip $Q_z$, contradicting the boundary condition. No other translation gives a real-valued phase for both wavevectors. 

The two hard-wall domains are therefore \textit{not} related by any allowed lattice translation. Geometric compatibility requires continuous interpolation of the antiphase tilt components across the wall, with local suppression of one or more components,
\begin{equation}
  Q_{x,y}(0) \to 0,
\end{equation}
producing genuine spatial gradients,
\begin{equation}
  \partial_z Q_{x,y} \neq 0.
\end{equation}
Crucially, because the c-flip preserves $Q_z$, only the $R_4^+$ components pass through zero, $Q_z$ remains large throughout the wall. The interpolation is therefore Néel-like: rather than collapsing in amplitude, the staggered Sr order parameter rotates continuously from one orientation to the other across the boundary, consistent with the experimental observation of continuous directional rotation and weak amplitude suppression at the hard boundary
(Fig.~\ref{fig: Long range Domain Mapping}c).

The consequences for the roto-flexoelectric coupling term (Eq.~\eqref{eq:roto-flexo-easy}) follow directly from this Néel character. The preserved $M_3^+$ component $Q_z \approx Q_z^0$
remains large at the wall centre precisely where the gradients $\partial_k Q_{x,y}$ are steepest. The coupling term $Q_z \cdot \partial_k Q_{x,y}$ therefore has a large prefactor exactly
where the gradient is maximised, the opposite situation to the easy wall, where Ising collapse suppresses $Q_j$ at the point of steepest gradient. Rather than counteracting the coupling, the preserved $M_3^+$ component amplifies it, driving the strong interfacial polarisation
observed experimentally at hard boundaries.

The Ising versus Néel interpolation character of the two boundary types is therefore not merely a phenomenological distinction but is mechanistically connected to the presence or absence of polarisation: it is a direct consequence of whether all tilt components reverse together (easy, Ising, polar response suppressed) or only the $R_4^+$ components pass through zero while $M_3^+$ is preserved (hard, Néel, polar response amplified). This connection between interpolation character
and polar response is specific to multicomponent $a^-a^-c^+$ orthorhombic perovskites and does not arise in single-component antiferrodistortive systems such as SrTiO$_3$ \cite{morozovskaInterfacialPolarizationPyroelectricity2012}, where no preserved component exists to amplify the gradient coupling.

\subsubsection*{Summary}

Both easy and hard APBs involve a reversal of the $R_4^+$ antiphase tilt components of $\mathbf{Q}$ and are therefore identified by the same discrete $180^\circ$ phase reversal of the staggered Sr modulation in the Fourier phase map.

\begin{figure}[hbt!]
	\centering
	\includegraphics[width=\columnwidth]{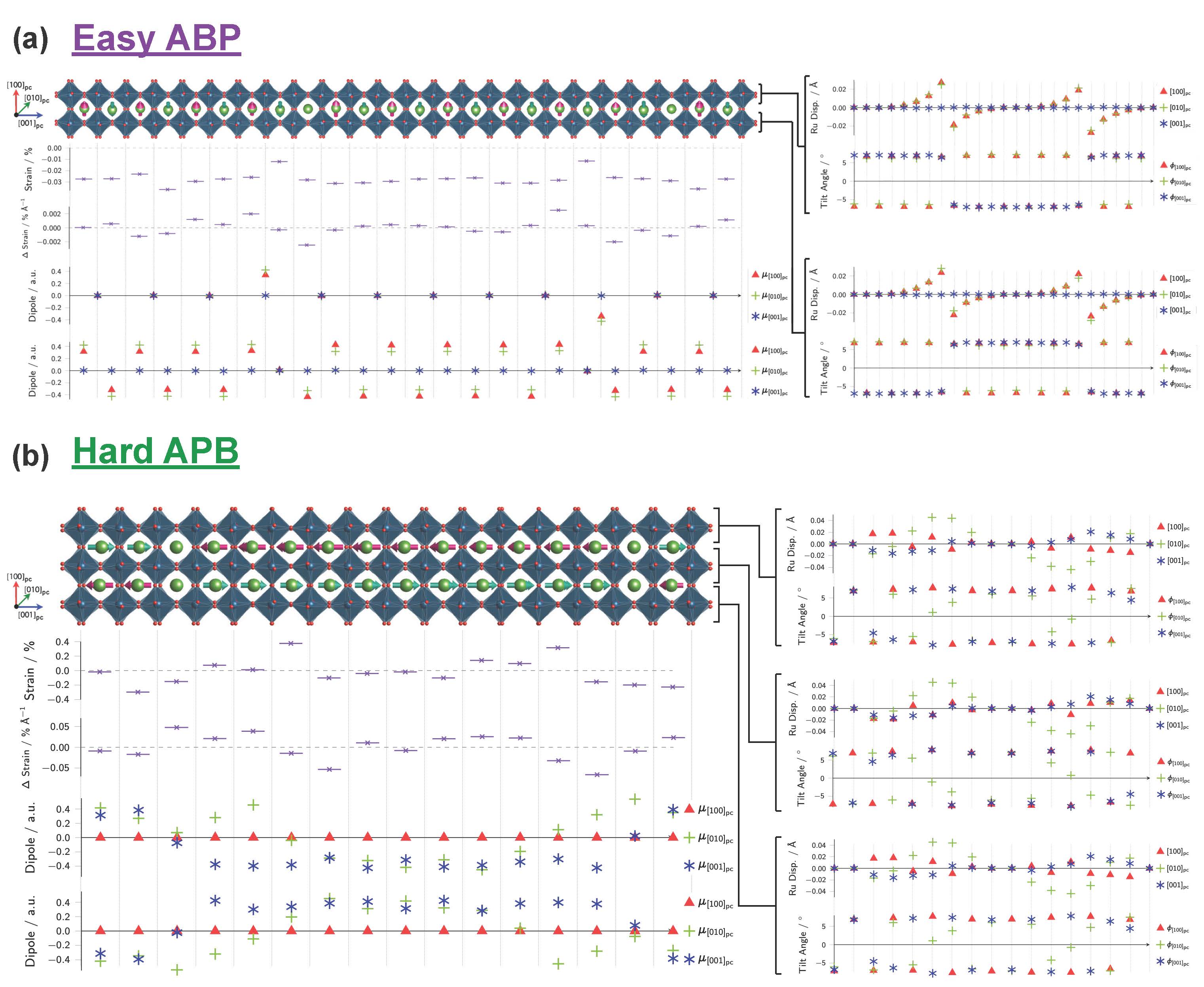}
\caption{\small
\textbf{Strain, polar displacement, and order-parameter profiles at easy and hard antiphase boundaries from first-principles calculations.}
\textbf{(a)} Easy antiphase boundary. Left panels, from top to bottom: spatially resolved longitudinal strain and strain gradient across the boundary; electric dipole moment of the full orthorhombic unit cell; dipole moment of the pseudocubic unit cell. Right panels: Ru off-centring displacement and octahedral tilt angle extracted independently from the top and bottom atomic rows of the supercell.
\textbf{(b)} Hard antiphase boundary. Left panels show the same quantities as in \textbf{(a)}. Right panels: Ru off-centring displacement and octahedral tilt angle extracted from the top, middle, and bottom atomic rows, resolving any depth-dependent variation in the polar and structural response across the supercell.}
\label{fig: Antiphase Boundary}
\end{figure}

At easy boundaries, the discrete translation $\mathbf{t}_4 = \tfrac{1}{2}[110]_\mathrm{pc}$ simultaneously reverses all three order-parameter components. The wall interpolation is therefore Ising-like: all components of $\mathbf{Q}$ collapse toward zero together at the wall centre without rotating. Within the roto-flexoelectric coupling term $P_i \sim \lambda_{ijkl}\,Q_j\,\partial_k Q_l$, the gradients $\partial_k Q_{x,y}$ and $\partial_k Q_z$ are largest at the wall centre, but all prefactors $Q_j$ are simultaneously suppressed toward zero by the Ising collapse. The coupling term is therefore strongly suppressed at the point where gradients are largest, and no net polarisation is generated. Whether this suppression is exact depends on the relative magnitudes of the rotostrictive tensor components $\lambda_{ijkl}$ and the tilt amplitudes, a quantitative question which the first-principles calculations presented in Supporting Information Fig.~\ref{fig: Antiphase Boundary} address.

At hard boundaries, the c-flip operation preserves $Q_z$ while reversing $(Q_x, Q_y)$, and no parent-cubic translation can reconcile this configuration. The wall interpolation is therefore N\'{e}el-like: $(Q_x, Q_y)$ pass through zero while $Q_z$ remains large throughout. The preserved $M_3^+$ component $Q_z \approx Q_z^0$ provides a large, non-zero prefactor in the coupling term precisely where the gradients $\partial_k Q_{x,y}$ are steepest, amplifying rather than suppressing the roto-flexoelectric response. This amplification drives the strong interfacial polarisation observed experimentally at hard boundaries.

The Ising versus N\'{e}el interpolation character is therefore not merely a phenomenological distinction but is the direct mechanistic origin of polar selectivity in $a^-a^-c^+$ orthorhombic perovskites. It arises specifically from the multicomponent nature of the order parameter: in single-component antiferrodistortive systems such as SrTiO$_3$~\cite{morozovskaInterfacialPolarizationPyroelectricity2012}, no preserved component exists to amplify the gradient coupling at hard boundaries, and the easy/hard polarisation distinction is less pronounced. The properties of both boundary classes are summarised in Table~\ref{tab:summary} of the main text.

\begin{table}[htbp]
\centering

\label{tab:summary_SI}
\begin{tabular}{lll}
\hline
Property & Easy boundary & Hard boundary \\
\hline
Domain relation    & Translation $\mathbf{t}_4 = \tfrac{1}{2}[110]_\mathrm{pc}$ & c-flip: no translation equivalent \\
$R_4^+$ sign change  & Yes (via $\mathbf{t}_4$) & Yes (via c-flip) \\
$M_3^+$ sign change  & Yes (via $\mathbf{t}_4$) & No (preserved by c-flip) \\
Wall interpolation   & Ising-like (all components $\to 0$) & N\'{e}el-like ($Q_z$ preserved) \\
Order-param.\ gradient & Present, but $Q_j \to 0$ at wall centre & Present, $Q_z^0$ amplifies coupling \\
Roto-flexo coupling  & Suppressed & Amplified \\
Polarisation        & Absent & Present \\
\hline
\end{tabular}

\caption{Summary of easy and hard antiphase boundary properties in
$a^-a^-c^+$ SrRuO\textsubscript{3}, reproduced from Table~\ref{tab:summary} 
of the main text for reference.}

\end{table}

\newpage

\subsection*{Supplemental Note C: 90$^\circ$ ferroelastic wall: 
symmetry origin of polar response}

The $90^\circ$ ferroelastic wall separates adjacent ferroelastic variants related by a $90^\circ$ reorientation of the orthorhombic axes within the plane of the membrane. The experimental observations presented in the main text (Fig.~\ref{fig: Short range Polarisation}e) identify these as boundaries between X/X$^*$ and Y/Y$^*$ variants, in which the orthorhombic long axis rotates by $90^\circ$ within the membrane plane. Additional $90^\circ$ boundaries involving the Z/Z$^*$ variant, in which the long axis rotates from in-plane to out-of-plane, are also present in the freestanding membrane but are not the subject of the present analysis. The domain relationship for the X-Y wall is generated by a $90^\circ$ rotation about the out-of-plane pseudocubic axis $\hat{z}_\mathrm{pc}$,
\begin{equation}
  \mathbf{R}_{z,90} =
  \begin{pmatrix} 0 & 1 & 0 \\ -1 & 0 & 0 \\ 0 & 0 & 1 
  \end{pmatrix},
  \qquad \det(\mathbf{R}_{z,90}) = +1,
  \label{eq:Rz90}
\end{equation}
which is a proper rotation. Unlike the c-flip operation at hard antiphase boundaries ($\det = -1$), no additional sign change arises from the axial-vector transformation law.

\subsubsection*{Transformation of the tilt order parameter at 
the X-Y wall}

Because octahedral rotations are axial vectors, the tilt order parameter transforms under $\mathbf{R}_{z,90}$ as
\begin{equation}
  \boldsymbol{\phi}' 
  = \det(\mathbf{R}_{z,90})\,\mathbf{R}_{z,90}\,\boldsymbol{\phi}
  = (+1)
    \begin{pmatrix} \phi_y \\ -\phi_x \\ \phi_z \end{pmatrix}
  = (\phi_y,\,-\phi_x,\,\phi_z).
  \label{eq:Rz90_transform}
\end{equation}
The three components therefore transform as follows:
\begin{itemize}
  \item $\phi_x \to \phi_y$ and $\phi_y \to -\phi_x$: the two antiphase $R_4^+$ components undergo a continuous in-plane rotation in the $(\phi_x, \phi_y)$ tilt plane. Crucially, neither component is required to reverse sign, so neither needs to pass through zero at the wall centre. The interpolation of $(\phi_x, \phi_y)$ across the wall is therefore entirely N\'{e}el-like in character: the tilt vector rotates continuously without amplitude suppression.
  \item $\phi_z \to \phi_z$: the in-phase $M_3^+$ component is \textit{strictly preserved} by the domain-relating operation. Because the X and Y variants share the same $\phi_z$ magnitude and sign, no discontinuity in $\phi_z$ exists across the wall and no suppression of this component is required at any point.
\end{itemize}

This is qualitatively distinct from the hard antiphase boundary, where the $R_4^+$ components must pass through zero (because they reverse sign), producing steeper gradients $\partial_k \phi_{x,y}$ at the wall centre. At the X-Y wall, the $R_4^+$ components rotate rather than collapse, so the spatial gradients $\partial_k \phi_{x,y}$ are distributed more gradually across the wall width. Nevertheless, genuine spatial gradients exist, $\partial_k \phi_{x,y} \neq 0$, and the preserved $M_3^+$ component $\phi_z \approx \phi_z^0$ remains large throughout, providing a non-zero prefactor in the roto-flexoelectric coupling term $\phi_z\,\partial_k\phi_{x,y}$ at every point across the wall.

\subsubsection*{Consequences for roto-flexoelectric coupling}

The roto-flexoelectric coupling term $P_i \sim \lambda_{ijkl}\,\phi_j\,\partial_k\phi_l$ is therefore non-zero throughout the X-Y wall. Unlike the easy antiphase boundary, where Ising collapse suppresses all $\phi_j$ precisely where gradients are steepest, and unlike the hard antiphase boundary, where the $R_4^+$ components vanish at the wall centre while $\phi_z$ amplifies the coupling, the X-Y wall presents a situation in which \textit{no component vanishes}: $\phi_z$ is preserved and $(\phi_x, \phi_y)$ rotate continuously. The roto-flexoelectric coupling is therefore active across the full width of the wall, though the absence of a sharp zero-crossing of $(\phi_x,\phi_y)$ means the peak gradient is lower than at a hard APB of comparable width. The quantitative contribution of each rotostrictive tensor component $\lambda_{ijkl}$ to the observed hydrostatic strain gradient requires a full tensor analysis that lies beyond the scope of the present work.

\subsubsection*{Elastic accommodation and unified flexoelectric 
picture}

The polar response at the $90^\circ$ wall is unified under the flexoelectric constitutive relation: polarisation couples to the observed hydrostatic strain gradient, which itself has two contributing sources. The first is the elastic accommodation of the deviatoric eigenstrain mismatch between variants. At the X-Y wall, the two short orthorhombic axes $a_o$ and $c_o$ are exchanged between in-plane directions, producing a deviatoric eigenstrain mismatch of magnitude $(c_o - a_o)/(\sqrt{2}\,a_\mathrm{pc}) \approx 0.6\%$ ($a_o = 5.514$~\AA, $c_o = 5.548$~\AA, $a_\mathrm{pc} \approx 3.93$~\AA). Although the far-field bulk mismatch between variants is purely deviatoric, elastic accommodation at the finite-width wall necessarily generates a strain field with both deviatoric and hydrostatic components, as required by elastic compatibility~\cite{Khachaturyan1991}. The second source is rotostriction: as the tilt amplitude interpolates through the wall, spatial variation of $\boldsymbol{\phi}$ generates an additional hydrostatic strain component through the rotostrictive coupling $F_\text{roto} = -\lambda_{ijkl}\,u_{ij}\,\phi_k\,\phi_l$ (Eq.~\ref{eq:rotostriction} of Supplemental Note~A). The spatially varying hydrostatic strain observed experimentally at the embedded walls (Fig.~\ref{fig: Short range Polarisation}d) is therefore an emergent quantity arising from both sources, and the present data do not distinguish their relative contributions. Its spatial gradient drives polar displacement through the flexoelectric constitutive relation (Eq.~\ref{eq:flexo_polar} of the main text).

This mechanism is expected to operate in any ferroelastic perovskite with a symmetry-lowering octahedral tilt, regardless of whether the tilt pattern is single- or multi-component, and is not specific to the $a^-a^-c^+$ symmetry of SrRuO$_3$.

The characteristic length scale of $\sim$4~nm observed experimentally reflects the spatial extent of the hydrostatic strain gradient field at each embedded $90^\circ$ wall, set by the wall width and the deviatoric eigenstrain mismatch, rather than by the electrostatic screening length alone; the latter is sub-unit-cell in SrRuO$_3$ \cite{BraicSciRep2015} and therefore does not limit the observed length scale.

Together, the easy and hard antiphase boundaries and the $90^\circ$ X-Y ferroelastic wall illustrate that polar response in multicomponent tilt systems is determined not simply by the presence of structural gradients, which exist at all three boundary types, but by the interplay between the symmetry of the domain-relating operation, the interpolation character of the tilt field, and the variant-dependent elastic 
properties of the adjacent domains.

\newpage
\subsection*{Supporting Figures}

\begin{figure}[hbt!]
	\centering
	\includegraphics[width=\columnwidth]{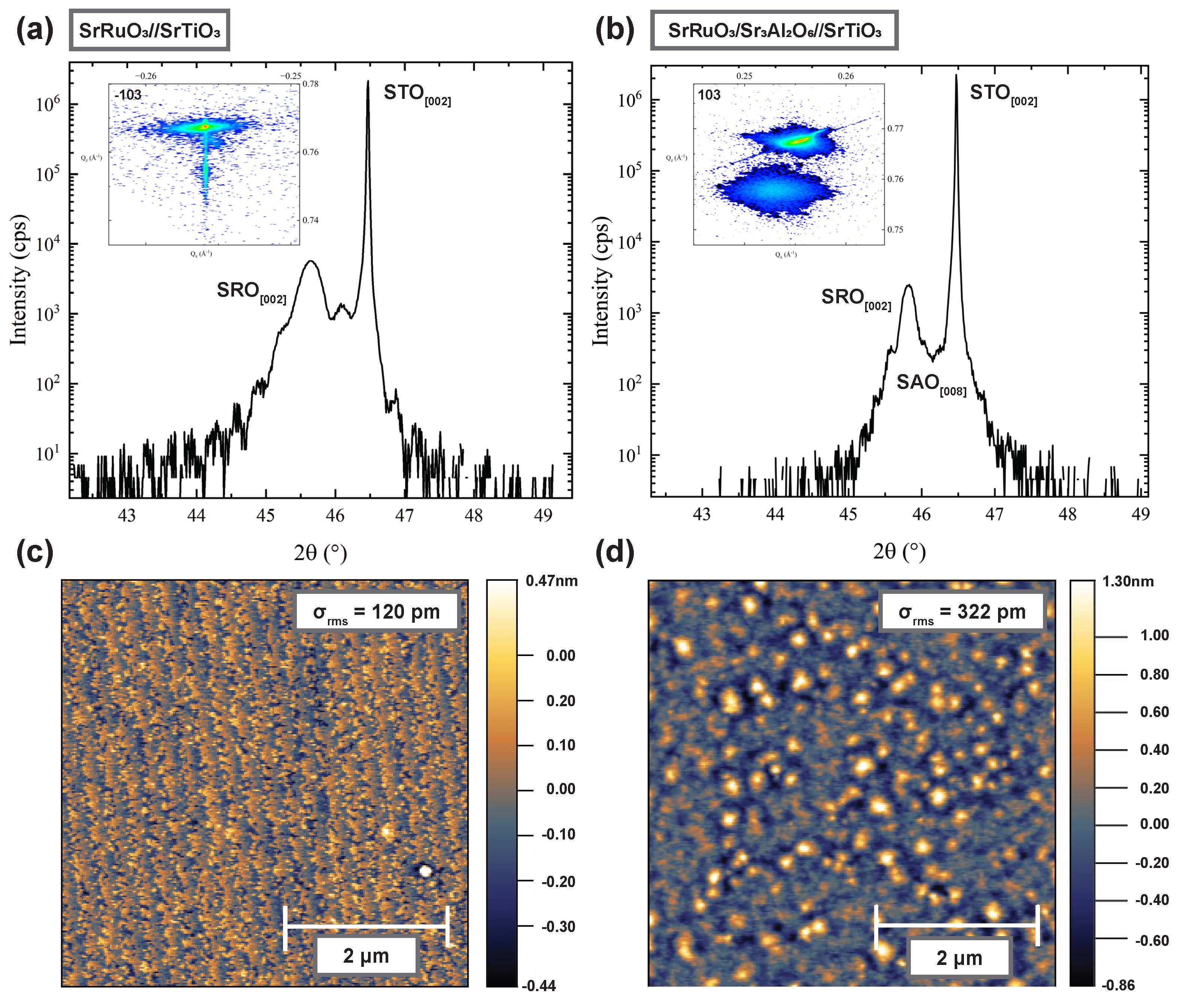}
    \caption{\small \textbf{Structural and morphological characterisation of epitaxial SrRuO$_3$ heterostructures.}
\textbf{(a)} Specular $\theta$/$2\theta$ X-ray diffraction scan of SrRuO$_3$/SrTiO$_3$, confirming phase-pure growth. Inset: reciprocal space map about the $\bar{1}03$ reflection, showing that the SrRuO$_3$ film peak is fully aligned with the SrTiO$_3$ substrate peak, confirming coherent epitaxial clamping.
\textbf{(b)} Specular $\theta$/$2\theta$ scan of the SrRuO$_3$/Sr$_3$Al$_2$O$_6$/SrTiO$_3$ heterostructure prior to membrane release. Inset: reciprocal space map about the $103$ reflection, showing lateral broadening of the SrRuO$_3$ film peak and a slight offset from the SrTiO$_3$ substrate peak, indicative of partial strain relaxation introduced by the Sr$_3$Al$_2$O$_6$ sacrificial layer.
\textbf{(c)} Atomic force microscopy surface topography of SrRuO$_3$/SrTiO$_3$ over a $5 \times 5\,\mu\text{m}^2$ area, revealing well-defined step-and-terrace structure with $\sigma_\mathrm{rms} = 120$~pm.
\textbf{(d)} Atomic force microscopy surface topography of SrRuO$_3$/Sr$_3$Al$_2$O$_6$/SrTiO$_3$ over a $5 \times 5\,\mu\text{m}^2$ area, showing increased surface roughness $\sigma_\mathrm{rms} = 322$~pm
consistent with the presence of the sacrificial layer.}

  \label{fig: All_XRD_AFM}
\end{figure}

\begin{figure}[hbt!]
	\centering
	\includegraphics[width=0.9\columnwidth]{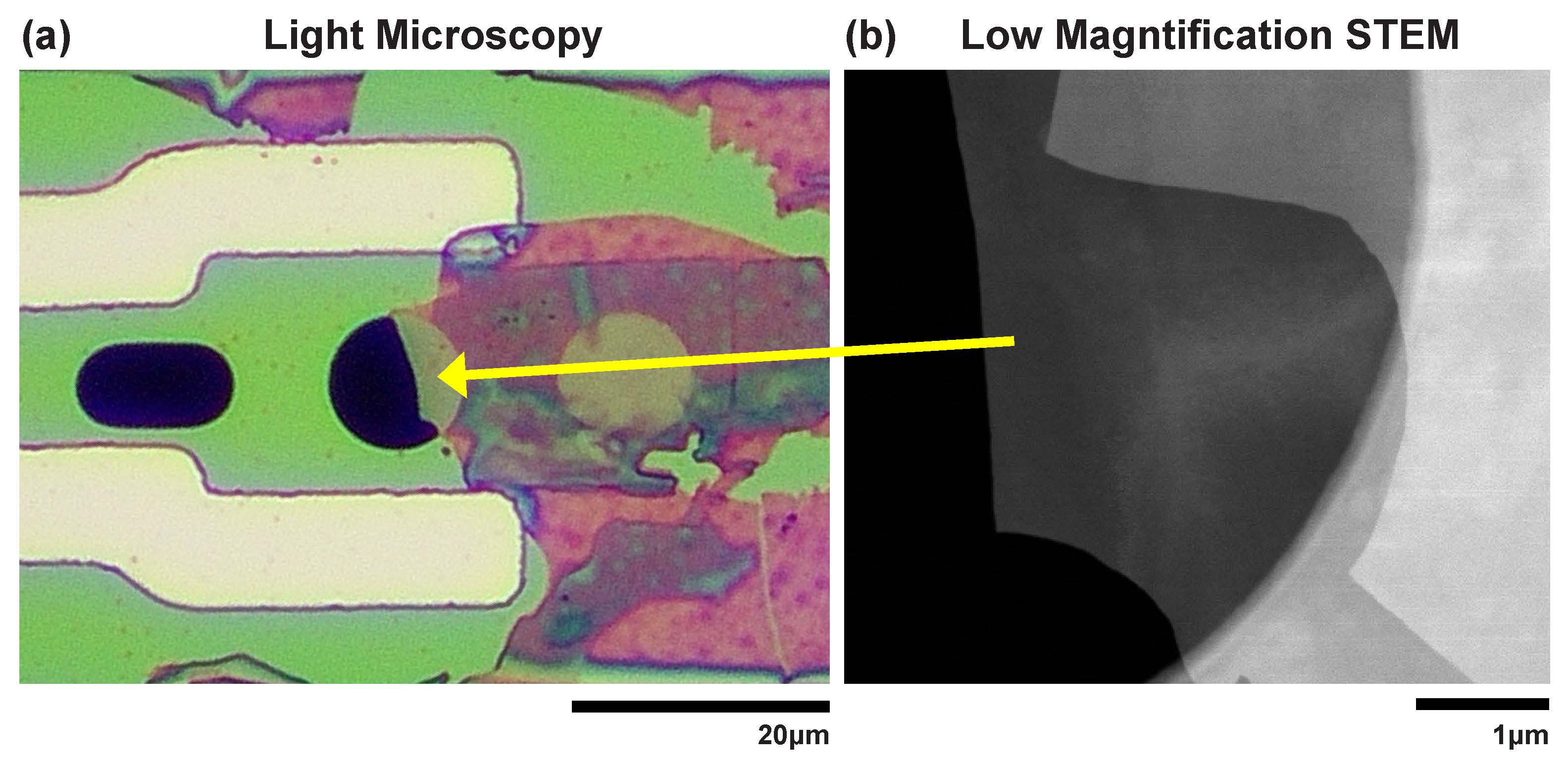}
\caption{\small \textbf{(a) Optical micrograph of the freestanding SrRuO$_3$ membrane transferred onto a TEM Nano-Chip and (b) low magnification STEM micrograph of the membrane} The SrRuO$_3$ membrane is visible suspended over the chip aperture.}
\label{fig: SRO_low_mag}
\end{figure}

\clearpage

\begin{figure}[hbt!]
	\centering
	\includegraphics[width=\columnwidth]{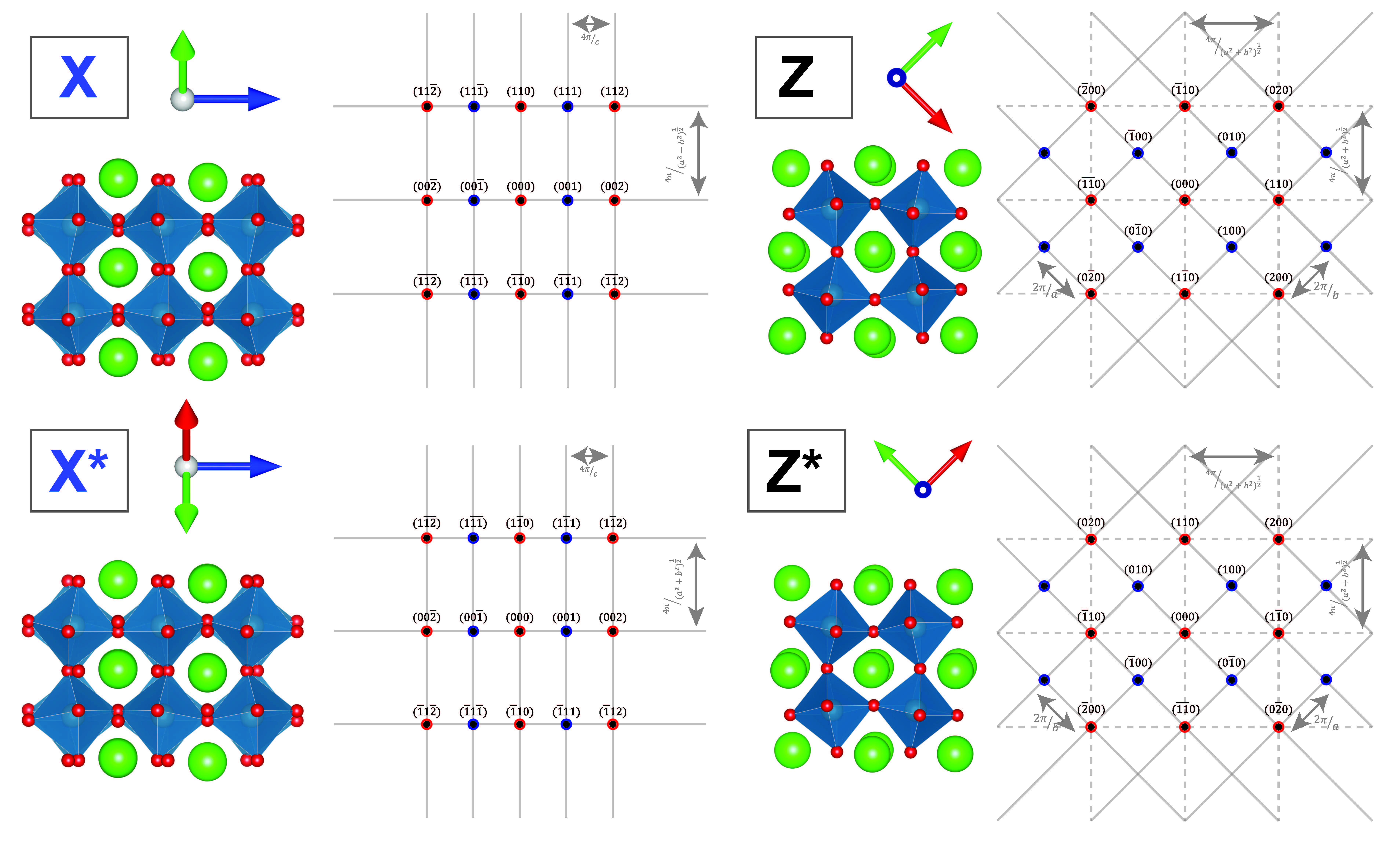}
\caption{\small \textbf{Crystallographic reference for SrRuO$_3$ ferroelastic variants and their diffraction signatures.}
Four panels showing the X, X$^*$, Z, and Z$^*$ orientation variants of orthorhombic SrRuO$_3$. Each panel displays the crystal structure with pseudocubic axes coloured as: $a$ (red), $b$ (green), $c$ (blue), alongside the corresponding simulated electron diffraction pattern. Spots outlined in red are reflections present in the parent pseudocubic structure; spots in blue are superlattice reflections arising from the symmetry lowering associated with the particular orthorhombic variant. The Z and Z$^*$ variants have the $c$-axis oriented out of plane, appearing cubic-like in projection. Y and Y$^*$ variants (not shown) are related to X and X$^*$ by a $90^\circ$ in-plane rotation and share the same diffraction signature with axes permuted accordingly. This figure serves as a crystallographic reference for the variant identification and FFT indexing used in the PLD analysis of the main text.}
  \label{fig: Table_for_Stuctures_and_HKL}
\end{figure}

\begin{figure}[hbt!]
	\centering
	\includegraphics[width=\columnwidth]{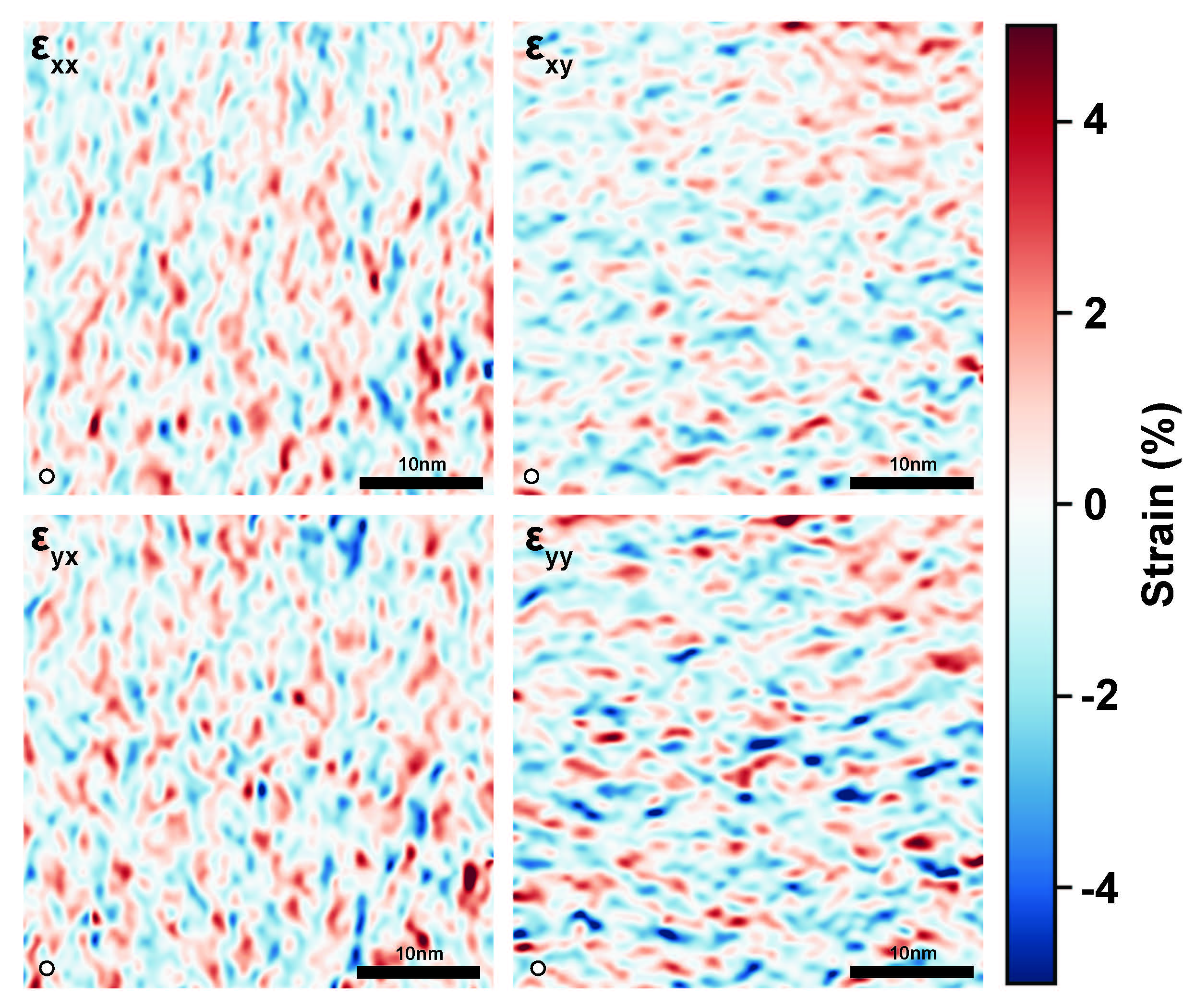}
\caption{\small \textbf{Strain field components extracted by Fourier phase
lock-in analysis.}
Spatially resolved $\varepsilon_{xx}$, $\varepsilon_{yy}$, $\varepsilon_{xy}$
and $\varepsilon_{yx}$ strain maps derived from ADF-STEM data of the freestanding
SrRuO$_3$ membrane using the KEMSTEM framework 
\cite{schnitzerQuantitativeApproachesMultiscale2025}. Coarsening length 
is shown in the bottom left. The $\varepsilon_{xx}$ and $\varepsilon_{yy}$ 
components show contrasting sign at the embedded $90^\circ$ ferroelastic 
walls, confirming the expected deviatoric eigenstrain mismatch between 
X/X$^*$ and Y/Y$^*$ variants, consistent with the elastic accommodation 
mechanism described in Supplemental Note~C.}
\label{fig: Strain}
\end{figure}

\begin{figure}[hbt!]
	\centering
	\includegraphics[width=\columnwidth]{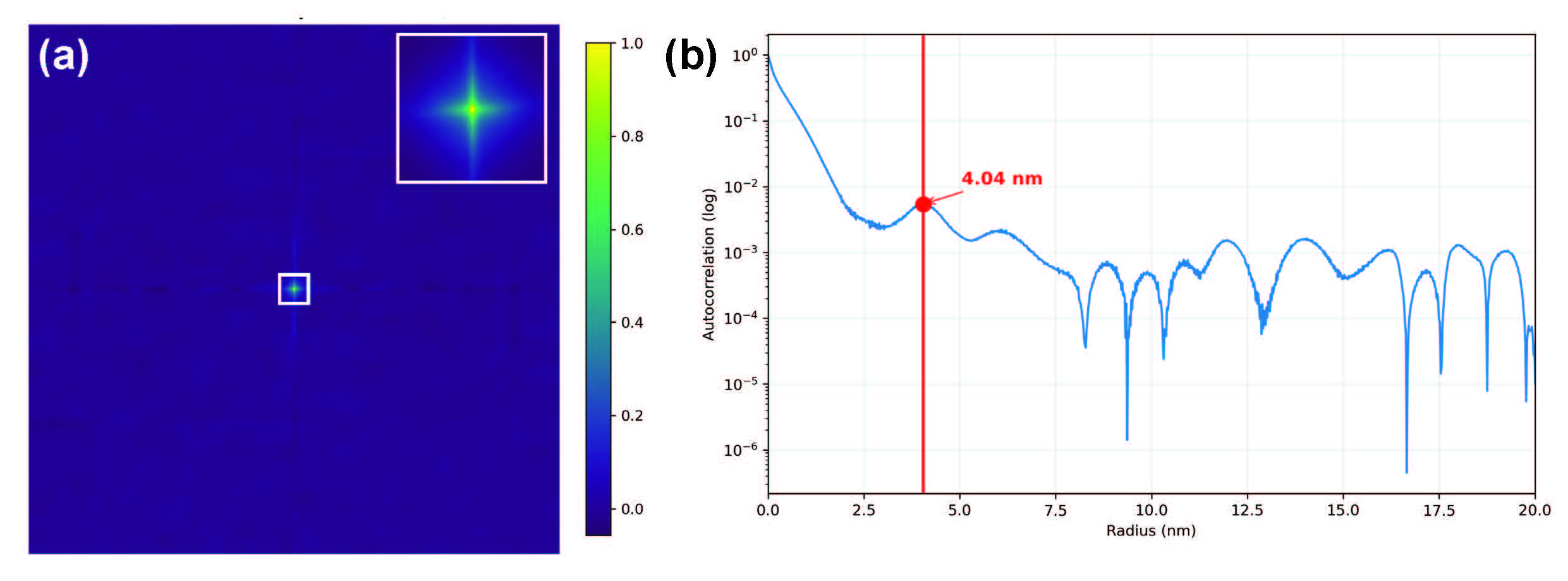}
  \caption{\small \textbf{Spatial autocorrelation analysis of the hydrostatic strain field in freestanding SrRuO$_3$.}
\textbf{(a)} Two-dimensional normalised autocorrelation of the hydrostatic strain map $\varepsilon_\mathrm{hyd}$, computed from mean-subtracted data with invalid regions masked prior to correlation. Inset: magnified view of the central region.
\textbf{(b)} Azimuthally averaged radial profile of the autocorrelation (log scale), with the first maximum at $r = 4.04$~nm marked. Radii below 2~nm are excluded to suppress the central autocorrelation peak. The characteristic length of $\approx 4$~nm defines the mean spatial periodicity of the hydrostatic strain modulation, setting the characteristic scale of the polar texture observed in Fig.~\ref{fig: Short range Polarisation}c.}
  \label{fig: Autocorrelation}
\end{figure}

\begin{figure}[hbt!]
	\centering
	\includegraphics[width=\columnwidth]{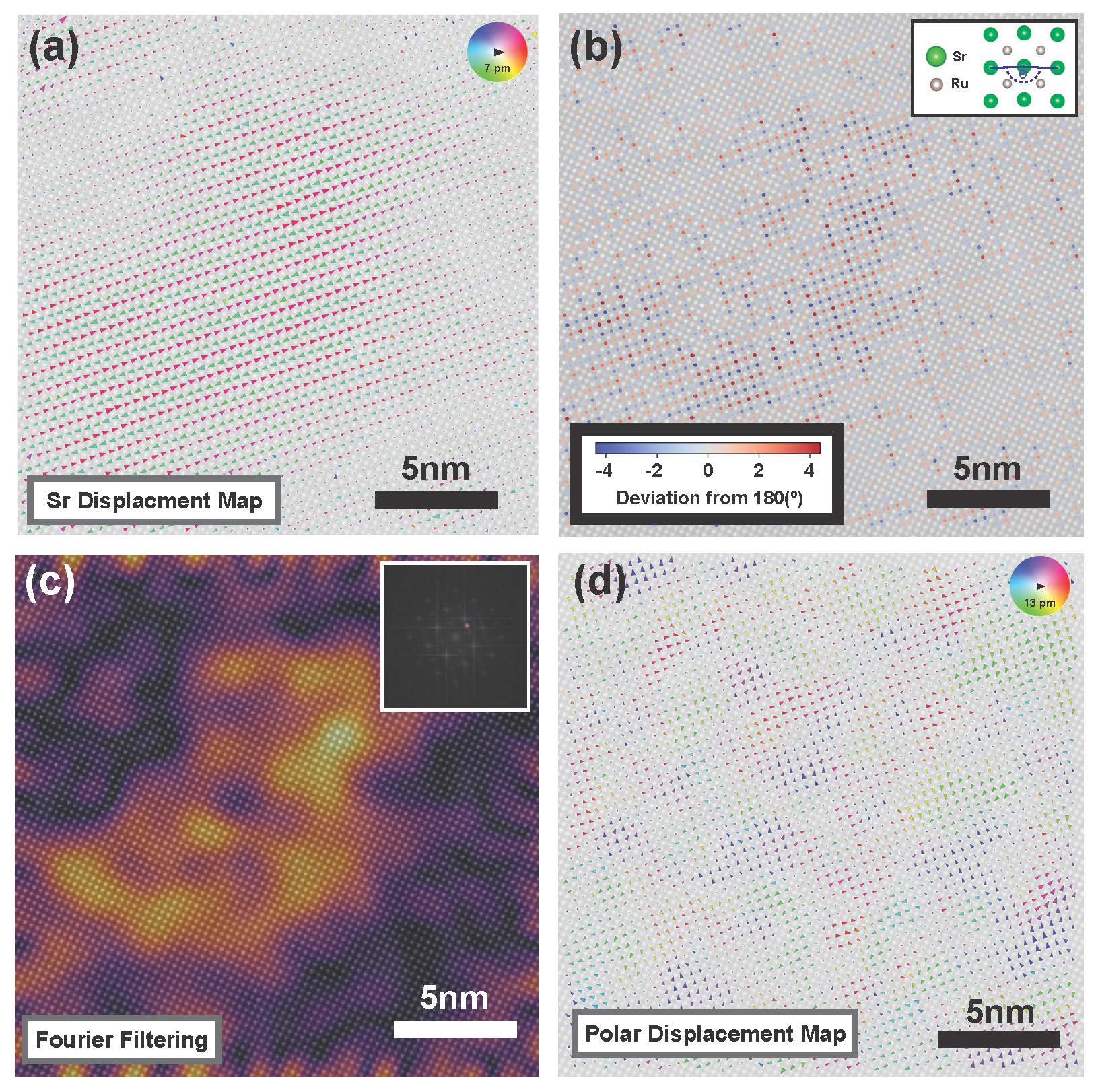}
    \caption{\small \textbf{Comparison of PLD and site-specific structural analysis in a nominally single Y/Y$^*$ domain of freestanding SrRuO$_3$.}
All panels show the same high-magnification region containing a central Y/Y$^*$ domain with a Z/Z$^*$ region at the periphery.
\textbf{(a)} PLD Sr displacement map of the region. The Y/Y$^*$ and Z/Z$^*$ domains are distinguishable, though domain boundaries appear broadened due to the non-local nature of the Fourier-based reconstruction.
\textbf{(b)} Site-specific map of Sr-Sr-Sr bond angle deviation from $180^\circ$ along the $[010]_\mathrm{pc}$ direction, yielding a more sharply defined central Y/Y$^*$ region and Z/Z$^*$ periphery. Spatial variations in bond angle directly reflect local distortions of the orthorhombic structure and are robust against projection artefacts, as this metric is computed from in-plane cation positions only.
\textbf{(c)} Fourier-filtered amplitude map of the $(11\bar{1})$ reflection, showing strong spatial correlation with the bond angle map in (b). Regions of suppressed bond angle modulation correspond to diminished Fourier amplitude, confirming that both metrics report consistently on the same underlying domain structure. The sharper boundary definition relative to (a) demonstrates the advantage of site-specific analysis for resolving embedded domain walls and short-range orthorhombic distortions at small length scales. \textbf{(d)} Polar displacement map, showing the same heterogeneous nanoscale polar texture observed in the main text (Fig.~\ref{fig: Short range Polarisation}c), here resolved at higher magnification. The polar displacements follow the perturbations in structure.}
  \label{fig: PLD_vs_Site_Analysis}
\end{figure}

\begin{figure}[hbt!]
	\centering
	\includegraphics[width=\columnwidth]{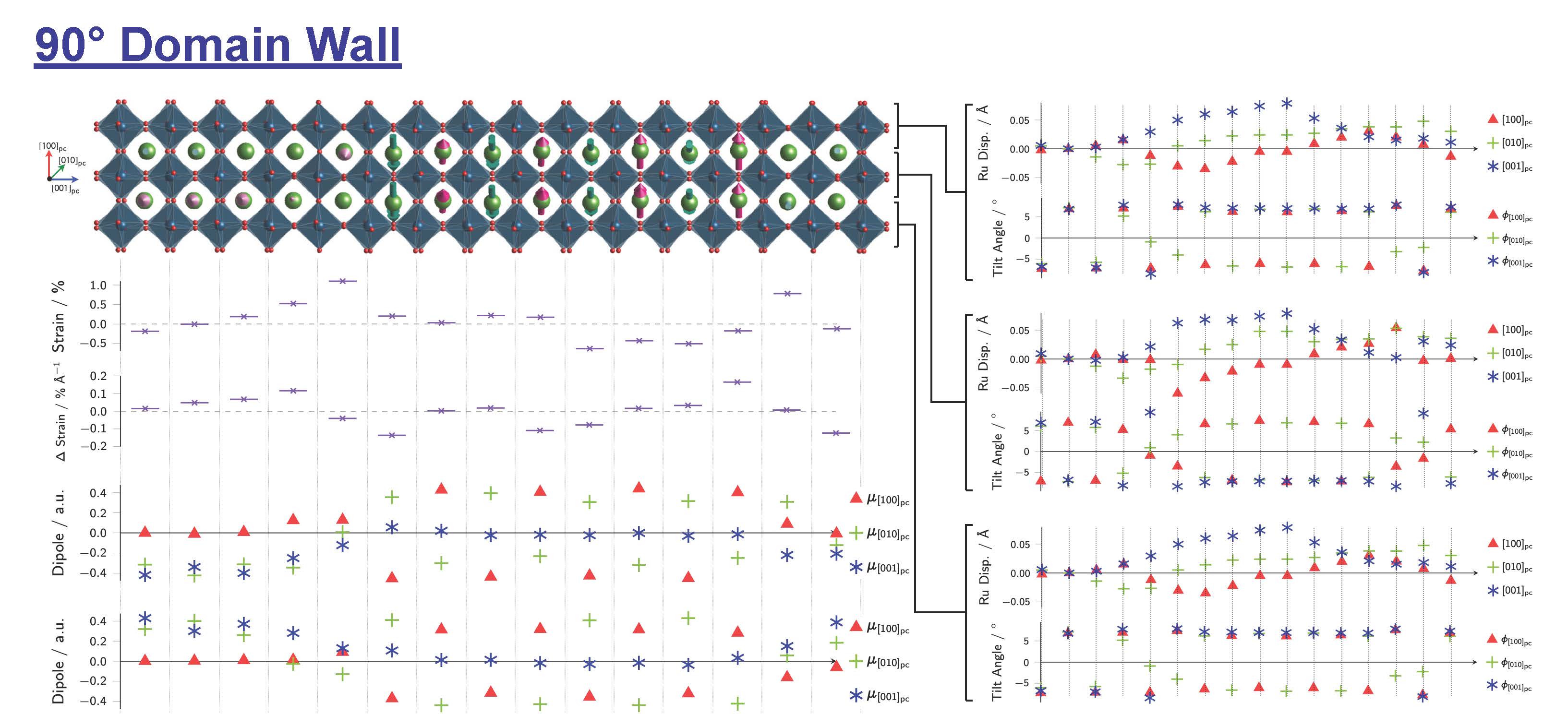}
\caption{\small
\textbf{Strain, polar displacement, and order-parameter profiles at the embedded 90$^\circ$ ferroelastic wall from first-principles calculations.}
Left panels, from top to bottom: spatially resolved longitudinal strain and strain gradient across the X-Y ferroelastic boundary; electric dipole moment of the full orthorhombic unit cell; dipole moment of the pseudocubic unit cell. Right panels: Ru off-centring displacement and octahedral tilt angle extracted independently from the top, middle, and bottom atomic rows of the supercell.}
\label{fig: 90_Deg_DFT}
\end{figure}

\begin{figure}[hbt!]
	\centering
	\includegraphics[width=\columnwidth]{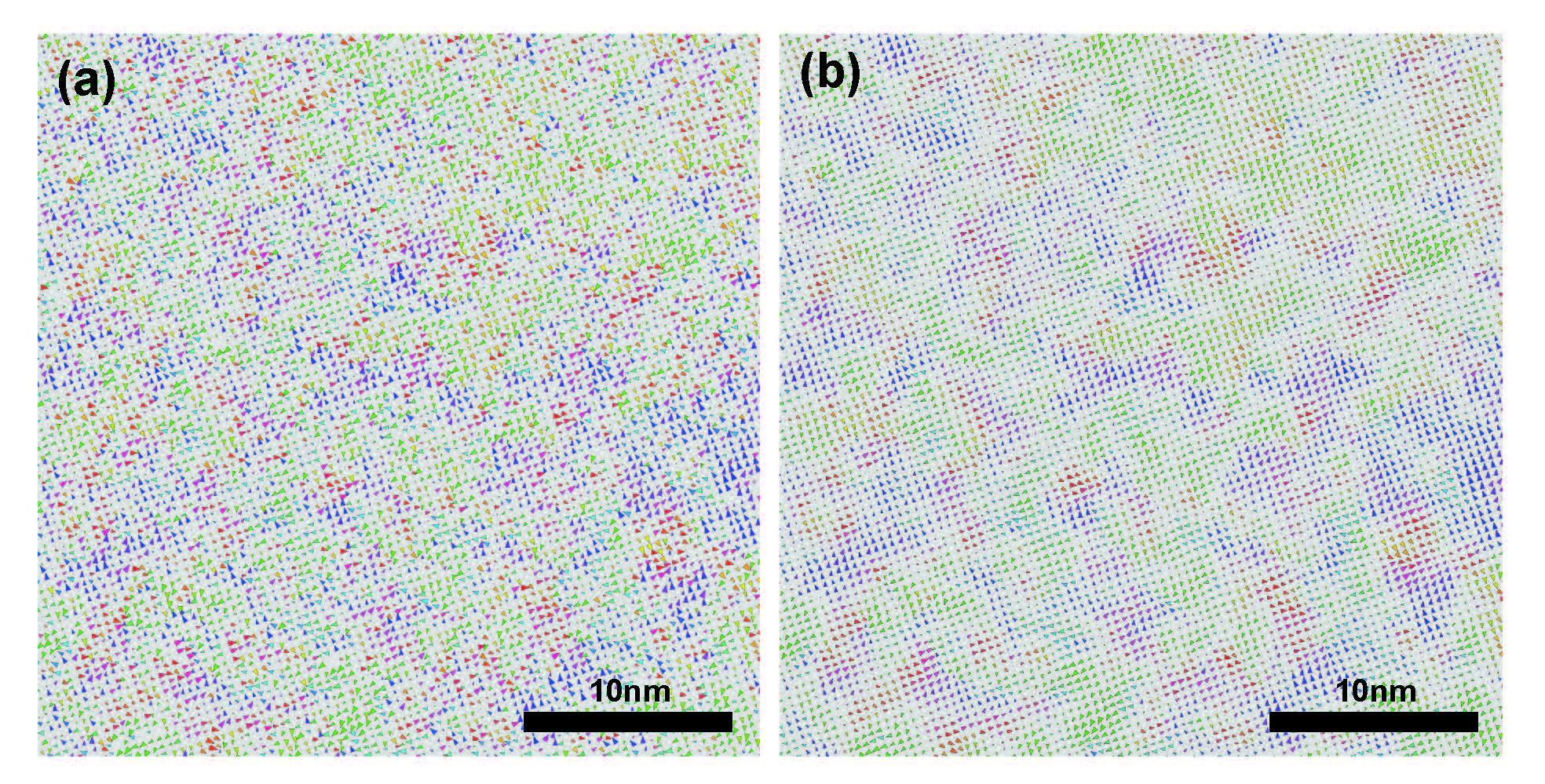}
\caption{\small \textbf{Effect of nearest-neighbour vector averaging on the polar displacement map.}
\textbf{(a)} Original polar displacement map showing Ru off-centring vectors at each unit cell, coloured by displacement direction.
\textbf{(b)} Smoothed polar displacement map obtained by replacing each site's displacement vector with the mean of all vectors within a 20-pixel search radius, subject to a minimum of 3 neighbours; in practice this averages each central site with approximately 4 surrounding neighbours, giving an effective smoothing kernel of $\sim$5 sites. Vector averaging is performed directly in Cartesian components. The smoothed map preserves the principal features of the polar texture, including the spatial distribution and characteristic orientation of polar domains, while reducing site-to-site noise, demonstrating that the nanoscale polar cluster structure reported in the main text is robust to local displacement fluctuations.}
  \label{fig: Smoothing}
\end{figure}

\begin{figure}[hbt!]
	\centering
	\includegraphics[width=0.5\columnwidth]{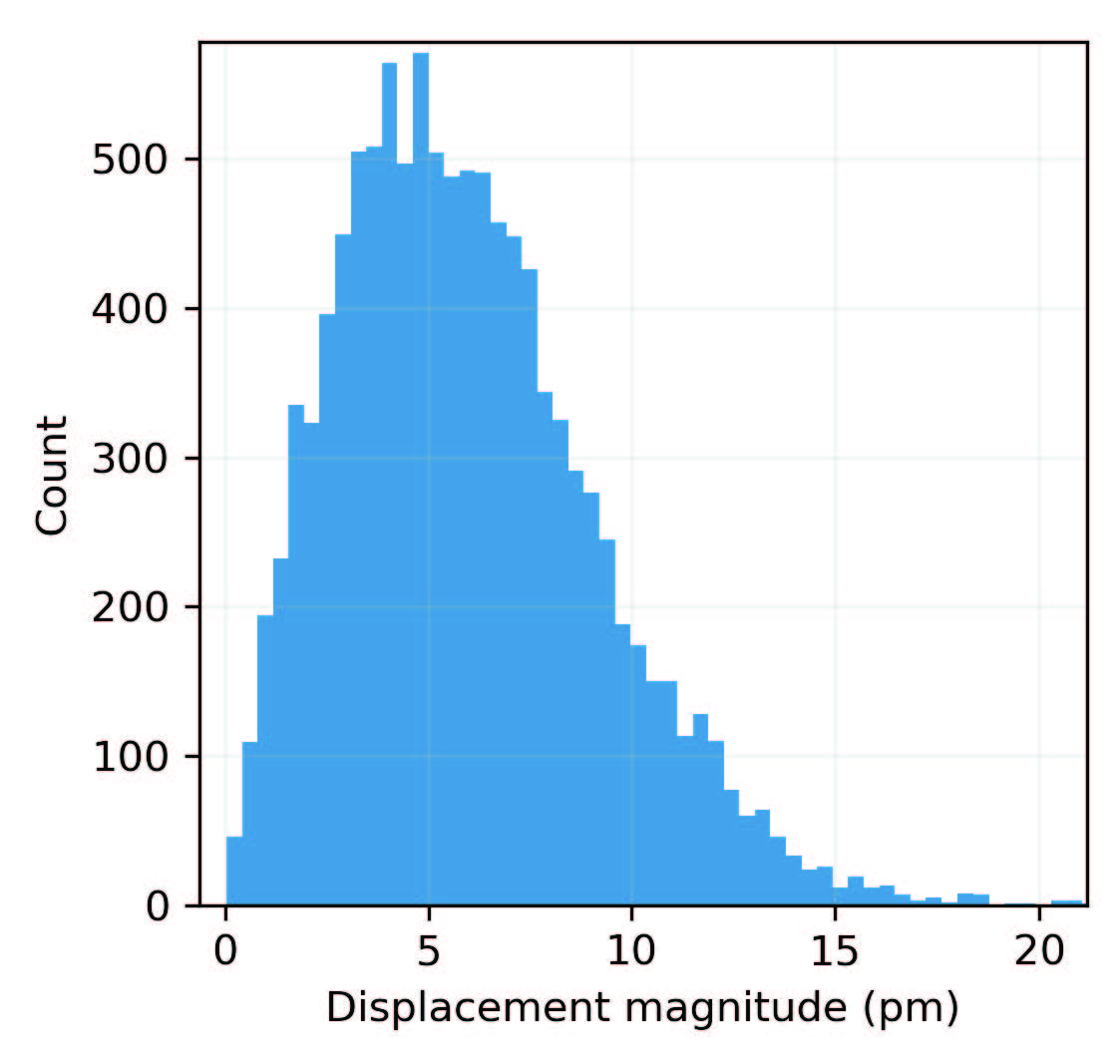}
\caption{\small \textbf{Statistical distribution of polar displacement magnitudes across the freestanding SrRuO$_3$ membrane.} 
Histogram of Ru off-centring displacement magnitudes extracted from ADF-STEM data ($N = 11245$ unit cells). The distribution has a mean displacement of 6.1~pm and a median of 5.6~pm, with a tail extending to displacements exceeding 13~pm at translation-inequivalent antiphase boundaries.}
  \label{fig: Histogram}
\end{figure}

\clearpage
\printbibliography

\newrefsection

\end{document}